\documentclass[article]{jfm}
\usepackage{amsmath,amssymb,amsbsy,graphicx,subfigure}
\usepackage[a4paper, left=2cm, right=2cm, top=2cm, bottom=2cm]{geometry}
\usepackage{natbib}

\pdfoutput=1

\newcommand{\diff}[3][]{\dfrac{\mathrm{d}^{#1}#2}{\mathrm{d}{#3}^{#1}}}

\newcommand{\pdiff}[3][]{\dfrac{\partial^{#1} #2}{\partial {#3}^{#1}}}

\title[Coalescence in a viscous fluid: Interface formation model]
{The coalescence of liquid drops in a viscous fluid: Interface
formation model}

\author[J.E. Sprittles and Y.D. Shikhmurzaev]
{J\ls A\ls M\ls E\ls S\ns E.\ns S\ls P\ls R\ls I\ls T\ls T\ls L\ls
E\ls S\footnote{E-mail: J.E.Sprittles@warwick.ac.uk} \and Y\ls U\ls
L\ls I\ls I\ns D.\ns S\ls H\ls I\ls K\ls H\ls M\ls U\ls R\ls Z\ls
A\ls E\ls V\footnote{E-mail: Y.D.Shikhmurzaev@bham.ac.uk}}

\affiliation{Mathematics Institute, University of Warwick, Coventry, CV4 7AL, UK, \newline School of Mathematics, University of
Birmingham, Birmingham B15 2TT, UK.}

\begin{document}

\label{firstpage} \maketitle

\begin{abstract}
The interface formation model is applied to describe the initial
stages of the coalescence of two liquid drops in the presence of a viscous
ambient fluid whose dynamics is fully accounted for. Our focus is on understanding (a) how this model's predictions differ from those of the conventionally used one, (b) what influence the ambient fluid has on the evolution of the shape of the coalescing drops and (c) the coupling of the intrinsic dynamics of coalescence and that of the ambient fluid. The key feature of the interface formation model in its
application to the coalescence phenomenon is that it removes the
singularity inherent in the conventional model at the onset of coalescence and describes the
part of the free surface `trapped' between the coalescing volumes as
they are pressed against each other as a rapidly disappearing
`internal interface'. Considering the simplest possible formulation of this model, we find experimentally-verifiable differences with the predictions of the conventional model showing, in particular, the effect of drop size on the coalescence process. According
to the new model, for small drops a non-monotone time-dependence of
the bridge expansion speed is a feature that could be looked for
in further experimental studies. Finally, the results of both models are compared to recently available
experimental data on the evolution of the liquid bridge connecting
the coalescing drops, and the interface formation model is seen to give a better agreement with the data.
\end{abstract}

\section{Introduction}

Coalescence, which is the process of two liquid volumes merging into
one, is central to numerous natural phenomena and a variety of
technological applications of fluids \citep{bellehumeur04,dreher99,grissom81,kovetz69}. In order to develop
these technologies, it is necessary to have a mathematical model of
the process which would allow one to reliably describe its dynamics
and hence minimize the time and resources on experimentation. The
current trend towards miniaturization of the fluid volumes
undergoing coalescence in various applications, e.g.\ in
biotechnologies \citep{quake05,seeman12} and additive manufacturing \citep{derby10,singh10},
makes it vital to accurately model the initial stages of
coalescence for which a mathematically singular description \citep{hopper84,richardson92,sprittles_pof2,sprittles_paper1}
would not be acceptable. The latter has been reinforced by recent
experiments \citep{paulsen11}, which employed a new experimental
technique that made it possible to probe the coalescence dynamics on
the `microfluidic' spatio-temporal scales inaccessible to traditional optical
methods used so far \citep{thoroddsen05,aarts05,wu04}.

The experimental breakthrough achieved in \cite{paulsen11} made it possible to test various
mathematical models of the process, and the first one to be tested
was the `conventional' model used in most studies
\citep{eggers99,duchemin03}. This model assumes that coalescence
takes place on a length scale below that of continuum mechanics, so
that, from the viewpoint of continuum mechanics, at the very onset
of the process one already has a single body of fluid consisting of
the two volumes that were brought into contact and a smooth, albeit
infinitesimal, liquid bridge connecting them. This scheme implies an
intrinsic singularity at the start of the process
\citep{hopper84,hopper90,hopper93a,hopper93b,richardson92} and, as
was shown in a numerical study that considered the conventional
model in its entirety \citep{sprittles_pof2}, it fails to describe
the newly available experimental data: the model strongly
overpredicts the speed at which the bridge connecting the coalescing
volumes expands. The situation was not remedied even when, for the
first time, the dynamics of the viscous gas surrounding the
coalescing volumes was fully accounted for \citep{sprittles_paper1}.
This makes it worthwhile examining how the conventional model could
be generalised to incorporate some additional physics that would
allow one to describe the coalescence process in a singularity-free
way.

\begin{figure}
     \centering
\includegraphics[scale=0.8]{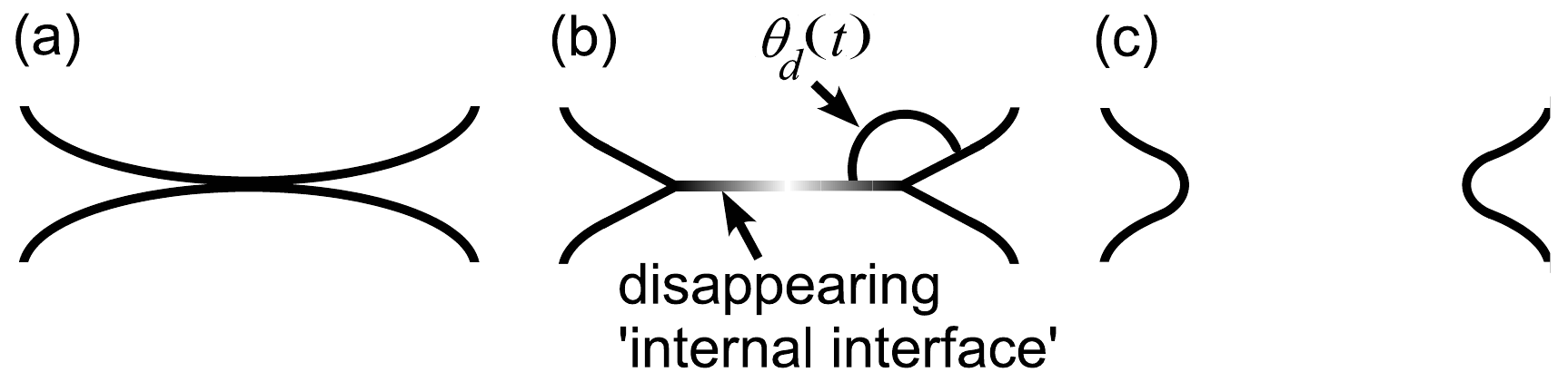}
\caption{\label{F:ifm_sketch} Sketch illustrating the scheme used in the interface formation/disappeance theory: the initial contact point (a) is followed by a fraction of the free surface being `trapped' between the bulk phases, forming a gradually disappearing `internal interface' (b), and, as the `internal interface' disappears and the `contact angle' $\theta_d$, being initially equal to $180^\circ$, relaxes to its `equilibrium' value of $90^\circ$, the conventional mechanism takes over (c).  The interface formation/disappearance model provides boundary conditions on interfaces, which are modelled as zero-thickness `surface phases'; these interfaces, including the `internal interface' in (b), are shown as finite-width layers for graphical purposes only.}
\end{figure}

A generalisation of the conventional model considered in
\cite{sprittles_pof2} is known as the interface formation model
\citep{shik07}. In the case of coalescence, this model suggests
that, as two liquid volumes are pressed against each other, a part
of the free surface becomes `trapped' between them, forming an
`internal interface' (Figure~\ref{F:ifm_sketch}). This interface gradually (although, in
physical terms, very quickly) disappears, losing its specific
`surface' properties, such as the surface tension, as the fluid
particles forming this interface adjust to their new environment and
turn into `ordinary' bulk particles. When the disappearance of the
internal interface separating the two volumes is complete, the
coalescence as such is over, one has a single body of fluid with a
{\it finite}-size smooth bridge connecting the initial volumes, and the
conventional model can take over without giving rise to any
mathematical singularities associated with the `smooth but infinitesimal' bridge it uses to start the process. Notably, before the internal interface
disappears, the residual surface tension associated with it can
sustain a (gradually disappearing) angle in the free surface, and,
once the coalescence is complete and the residual surface tension is
gone, the free surface becomes smooth, as required by the
conventional model which from that moment onwards can take over.

Simplifications outlined in \S\ref{S:simpler} will allow us to consider the ambient fluid to be either a gas or a second immiscible liquid. Notably, it has already been shown that an ambient gas affects the flow described using the conventional model \citep{sprittles_paper1}, even at
surprisingly small gas-to-liquid viscosity ratios, as the presence of the gas can qualitatively change the free-surface evolution in the early stages of the process, in particular, suppressing the formation of a toroidal bubble anticipated by earlier studies \citep{oguz89,duchemin03}, where the ambient gas was regarded as inviscid and dynamically passive. Given that in the interface formation model there is a cusp formed when the volumes first touch, which then evolves into a corner, one may expect a different effect of the gas dynamics than in the case of the conventional model where the interface is assumed to be smooth immediately after the onset of the process.  This aspect will be investigated.

It has been shown \citep{sprittles_pof2} that the interface
formation model's predictions are in better agreement than the conventional model's with the
experimental data on the early stages of coalescence
\citep{paulsen11}, and this fact justifies its further
investigation.  In \cite{sprittles_pof2} only a direct comparison of theory with experimental data was provided, without any parametric study of the model, which will be rectified here by (a) investigating how the surface variables evolve and depend on the constants characterizing material properties of the liquid-fluid system, (b) considering the effect of the ambient fluid on the coalescence process and (c) determining the coupling between the dynamics of the interfaces and that of the ambient fluid.  The main emphasis throughout will be on highlighting the specific features of the coalescence process, as described by the interface formation model, which distinguish it from the conventional model, with experimental-verification of these effects in mind. Only once this has been achieved will we consider a comparison to the experimental data with the effect of the ambient gas now fully accounted for.

\section{Problem formulation}\label{S:problem}

\begin{figure}
     \centering
\includegraphics[scale=0.6]{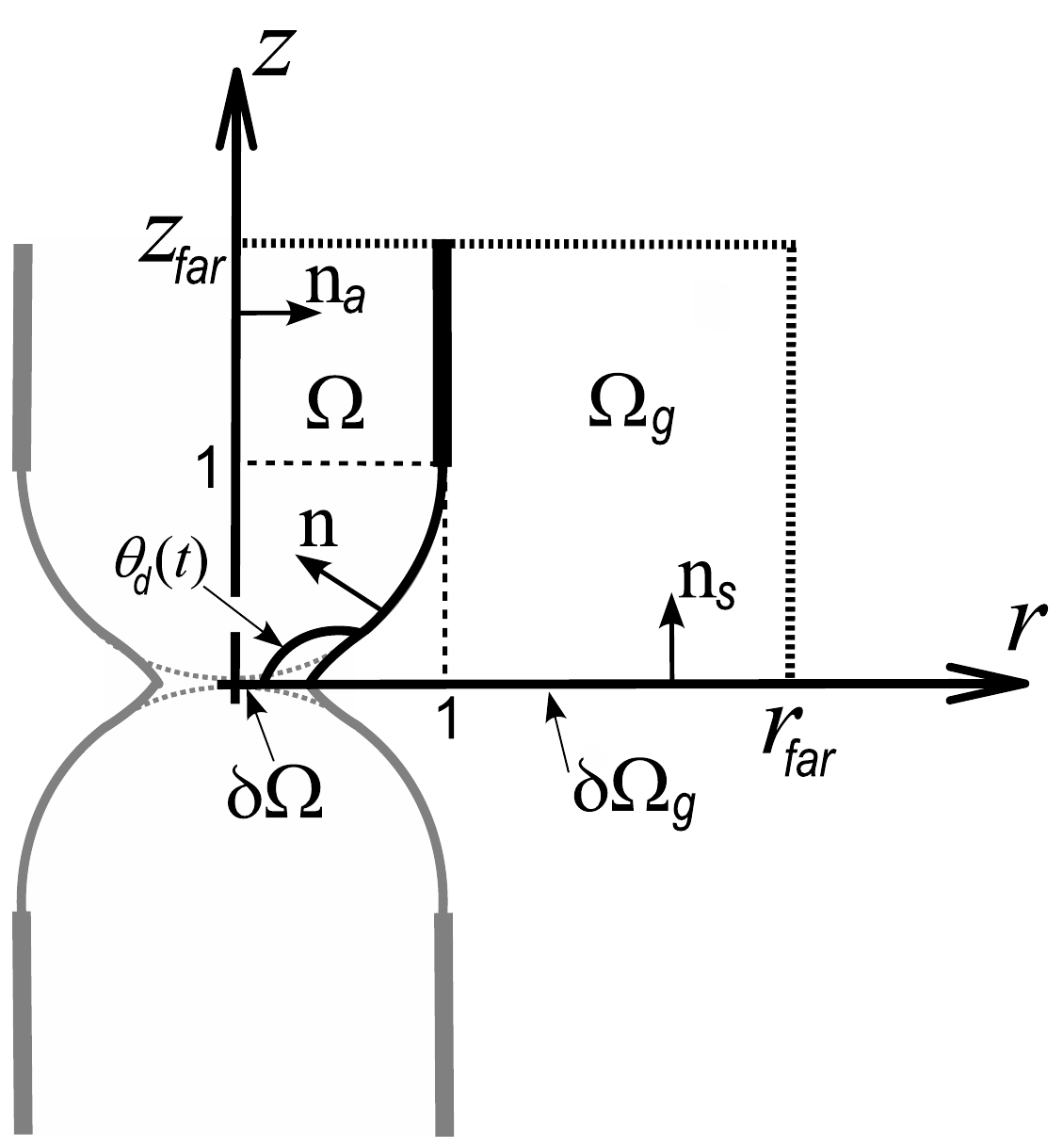}
 \caption{A sketch indicating aspects of the problem formulation for the coalescence of two identical hemispheres grown from syringes.}
 \label{F:sketch}
\end{figure}

Consider the axisymmetric coalescence of two drops that are grown
from two syringes and start coalescing when each of them reaches the
shape of a hemisphere (Fig.~\ref{F:sketch}). Both the liquid forming the
drops and the ambient fluid, whose dynamics we will also take into
account, will be described as incompressible Newtonian fluids with
constant densities $\rho$, $\rho_g$ and viscosities $\mu$, $\mu_g$,
respectively. The problem will initially be formulated for the case in which the ambient fluid is a gas, hence the subscripts `g', but we will later see that under certain simplifying assumptions, this formulation is equivalent to the case in which the ambient fluid is a second immiscible liquid.

The problem has an axial symmetry and symmetry
with respect to the plane tangential to the drops at the moment of
their initial contact, so that it is sufficient and convenient to
consider the flow only in the first quadrant of the $(r,z)$-plane of
the suitably chosen cylindrical coordinate system (Fig.~\ref{F:sketch})
and use the appropriate symmetry conditions on the axis and the
plane of symmetry. Using the initial radius of the drops $R$ as the
characteristic length scale, $U_v=\sigma_{1e}/\mu$, where $\sigma_{1e}$ is
the equilibrium surface tension of the fluid-liquid interface, as the
scale for velocities, $T_v=R/U_v=\mu R/\sigma_{1e}$ as the time scale,
and $\sigma_{1e}/R$ as the scale for pressure, we have that the (dimensionless) bulk
velocities $\mathbf{u}$, $\mathbf{u}_g$ and pressures $p$, $p_g$ in
the liquid and the ambient fluid satisfy the Navier-Stokes equations which in
the dimensionless form are given by
\begin{equation}\label{Navier-Stokes}
\nabla\cdot\mathbf{u} = 0,\qquad Re~\left[\pdiff{\mathbf{u}}{t} +
\mathbf{u}\cdot\nabla\mathbf{u}\right] =
\nabla\cdot\mathbf{P};\qquad \mathbf{P} = -p\mathbf{I} +
\left[\nabla\mathbf{u}+
 \left(\nabla\mathbf{u}\right)^T\right],
  \qquad \mathbf{r}\in\Omega
\end{equation}
\begin{equation}\label{Navier-Stokes-1}
\nabla\cdot\mathbf{u}_g = 0,\qquad
\bar{\rho}Re~\left[\pdiff{\mathbf{u}_g}{t} +
\mathbf{u}_g\cdot\nabla\mathbf{u}_g\right] =
\nabla\cdot\mathbf{P}_g;
 \qquad \mathbf{P}_g = -p_g\mathbf{I} +
\bar{\mu}\left[\nabla\mathbf{u}_g+\left(\nabla\mathbf{u}_g\right)^T\right],
 \qquad \mathbf{r}\in\Omega_g
\end{equation}
where $t$ is time; $\mathbf{P}$ and $\mathbf{P}_g$ are the stress
tensors in the liquid and the fluid, respectively; $\mathbf{I}$ is the
metric tensor of the coordinate system; $\Omega$ and $\Omega_g$
indicated the regions occupied by the liquid and the ambient fluid
(Fig.~\ref{F:sketch}). The non-dimensional parameters are the Reynolds
number $Re=\rho \sigma_{1e} R/\mu^2$ based on the liquid's properties,
the fluid-to-liquid density ratio $\bar{\rho}=\rho_g/\rho$ and the
corresponding viscosity ratio $\bar{\mu}=\mu_g/\mu$.

The interface formation model \citep{shik07}, which we will be using
to formulate the boundary conditions, states that part of the free
surface trapped between the drops ($\partial\Omega$ in
Fig.~\ref{F:sketch}) does not lose its specific surface properties, such
as the surface tension, instantly so that, until it does, one will have a
gradually disappearing `internal interface' whose (residual) surface
tension can sustain an angle in the free surface (Figure~\ref{F:ifm_sketch}). Pictorially, one
has a process analogous to dynamic wetting where the drops `spread'
over the separating plane of symmetry, with the `contact line' at
$r=r_c(t)$, $z=0$ leaving behind a gradually disappearing interface.
Using this analogy, the point in the $(r,z)$-plane at which the free
surface of the upper drop meets the plane of symmetry will be
referred to as the `contact line' and  the angle $\theta_d$ between
this free surface and the symmetry plane $z=0$ will be called the
`contact angle'. As the contact angle reaches its `equilibrium
value' of $90^\circ$, the free surface becomes smooth and the
conventional model takes over as the interface formation model simply reduces to it.

Both the free surface and the internal interface will be described
as two-dimensional `surface phases' characterized by their surface
tensions $\sigma_i$, surface densities $\rho^s_i$ and surface
velocities $\mathbf{v}^s_i$, where $i=1,2$, with subscripts 1 and 2
hereafter labelling the surface parameters of the free surface and
the internal interface, respectively. We will scale the surface
velocities with $U_v$, the surface tensions with $\sigma_{1e}$ and
the surface densities with a characteristic surface density
$\rho^s_{(0)}$.

On the free surface, besides the standard normal and tangential
stress boundary conditions,
\begin{equation}\label{norm-stress}
 \mathbf{n}\cdot\left(\mathbf{P}-\mathbf{P}_g\right)\cdot
 \mathbf{n} =\sigma_1 \nabla\cdot\mathbf{n},
\end{equation}
\begin{equation}\label{tang-stress-free}
 \mathbf{n}\cdot\left(\mathbf{P}-\mathbf{P}_g\right)\cdot
 (\mathbf{I}-\mathbf{n}\mathbf{n}) +\nabla\sigma_1=0,
\end{equation}
where $\mathbf{n}$ is a unit normal pointing into the liquid, one
has (a) the kinematic condition
\begin{equation}\label{kinem-free}
\pdiff{f}{t} + \mathbf{v}_1^s\cdot\nabla f = 0,
\end{equation}
where $f(r,z,t)=0$, with the a-priori unknown function $f$,
describes the evolution of the free-surface shape, and $\mathbf{v}_1^s$ is the corresponding velocity, (b) the surface
equation of state which in both interfaces will be taken in the
simplest linear form
\begin{equation}\label{eqstate}
\sigma_i=\lambda(1-\rho^s_i), \qquad (i=1,2),
\end{equation}
where $\lambda$ is a constant and $\rho^s_{i}$ is the dimensionless surface density, (c) the surface continuity equation
incorporating the mass exchange between the bulk and surface phases, and the corresponding equation for the normal component of the bulk
velocity
\begin{equation}\label{continuity-free}
 \epsilon\left[\frac{\partial\rho^s_1}{\partial t}
 +\nabla\cdot\left(\rho^s_1\mathbf{v}^s_1\right)\right]
 =-\left(\rho^s_1-\rho^s_{1e}\right),\qquad  (\mathbf{u}-\mathbf{v}^s_1)\cdot\mathbf{n}
 = Q(\rho^s_1-\rho^s_{1e}),
\end{equation}
where $\epsilon$, $\rho^s_{1e}$ and $Q$ are constants, (d) the
kinematic condition for the normal component of the ambient fluid velocity,
\begin{equation}\label{norm-gas-free}
 (\mathbf{u}_g-\mathbf{v}^s_1)\cdot\mathbf{n}=0,
\end{equation}
and (e) equations relating the tangential components of the bulk
velocities and stresses on the two sides of the interface, the surface
velocity and the gradients of surface tension
\begin{equation}\label{tang-bulk_vel-free}
 \hbox{$\frac{1}{2}$}
 \bar{\alpha}\mathbf{n}\cdot(\mathbf{P}+\mathbf{P}_g)\cdot(\mathbf{I}-\mathbf{n}\mathbf{n})
 =A(\mathbf{u}-\mathbf{u}_g)\cdot(\mathbf{I}-\mathbf{n}\mathbf{n}),
\end{equation}
\begin{equation}\label{Darcy-free}
\left[\mathbf{v}^s_1-\hbox{$\frac{1}{2}$}(\mathbf{u}+\mathbf{u}_g)
 -\bar{\alpha}\nabla\sigma_1\right]
 \cdot(\mathbf{I}-\mathbf{n}\mathbf{n})=0,
\end{equation}
where $\bar{\alpha}$ and $A$ are constants.  The constant $\bar{\beta}$, which has been introduced in previous works \citep{shik07}, is $\bar{\beta}=A\bar{\alpha}^{-1}$. The
non-dimensional constants appearing in
(\ref{kinem-free})--(\ref{Darcy-free}) incorporate the
corresponding material constants whose physical meaning and values
for some systems are described elsewhere \citep[see][]{shik07}.

The location of the internal interface is known, $z=0$, and hence
$\mathbf{v}^s_2\cdot\mathbf{n}_s=0$, where $\mathbf{n}_s$ is a unit
normal to the plane of symmetry (Fig.~\ref{F:sketch}). Then, on this
interface one has only the tangential stress condition
\begin{equation}\label{tang-stress-intl}
 \mathbf{n}_s\cdot\mathbf{P}\cdot
 (\mathbf{I}-\mathbf{n}_s\mathbf{n}_s) +\nabla\sigma_2=0,
\end{equation}
analogous to (\ref{tang-stress-free}); the surface continuity
equation together with the corresponding condition on the normal
component of the bulk velocity,
\begin{equation}\label{continuity-intl}
 \epsilon\left[\frac{\partial\rho^s_2}{\partial t}
 +\nabla\cdot\left(\rho^s_2\mathbf{v}^s_2\right)\right]
 =-(\rho^s_2-1),
 \qquad
 \mathbf{u}\cdot\mathbf{n}_s
 = Q(\rho^s_2-\rho^s_{2e}),
\end{equation}
analogous to (\ref{continuity-free}); and equation
\begin{equation}\label{Darcy-intl}
 [4A(\mathbf{v}^{s}_2-\mathbf{u})-\bar{\alpha}(1+4A)\nabla\sigma_2]\cdot(\mathbf{I}-\mathbf{n}_s\mathbf{n}_s)
 =0,
\end{equation}
which relates the difference between the tangential components of
the surface and bulk velocity to the surface tension gradient in the
surface phase.

At the moving `contact line', $r=r_c(t)$, $z=0$, we have the
conditions of continuity of the surface mass flux and the force
balance in the projection on the symmetry plane,
\begin{equation}\label{cl1}
 \rho^s_1 (\mathbf{v}^s_1
 -\mathbf{U}_c)\cdot\mathbf{m}_1
 +\rho^s_2 (\mathbf{v}^s_2
 -\mathbf{U}_c)\cdot\mathbf{m}_2=0,
\end{equation}
\begin{equation}\label{cl2}
 \sigma_{2}+\sigma_{1}\cos\theta_d=0,
\end{equation}
where $\mathbf{U}_c=d\mathbf{r}_c/dt$; the unit vectors $\mathbf{m}_i$  are
normal to the contact line and inwardly tangential to the
free surface ($i=1$) and the plane of symmetry ($i=2$); $\theta_d$
is the `contact angle' (Fig.~\ref{F:sketch}). Equation (\ref{cl2}) is
analogous to the well-known Young's equation \citep{young05} that
introduces and determines the contact angle in the process of
dynamic wetting. Notably, the surface continuity equation
(\ref{continuity-intl}) together with (\ref{eqstate}) and
(\ref{cl2}) ensure that the completion of the coalescence process
associated with the internal interface reaching its equilibrium
state ($\rho^s_2=1$) results in the disappearance of this interface
($\sigma_2=0$) and the restoration of the familiar smooth free
surface ($\theta_d=90^\circ$) thus allowing the conventional model
to take over.

For the ambient fluid phase, on the plane of symmetry $z=0$ one has conditions
of impermeability and zero tangential stress
\begin{equation}\label{csym}
 \mathbf{u}_g\cdot\mathbf{n}_s=0,
 \quad
 \mathbf{n}_s\cdot\mathbf{P}_g\cdot(\mathbf{I}-\mathbf{n}_s\mathbf{n}_s)=\mathbf{0},
 \qquad (\mathbf{r}\in\partial\Omega_g),
\end{equation}
and, for the liquid phase, at the axis of symmetry $r=0$ one has the appropriate symmetry
conditions
\begin{equation}\label{axis}
\mathbf{u}\cdot\mathbf{n}_a = 0,
 \quad
 \frac{\partial}{\partial r}
 [ \mathbf{u} \cdot (\mathbf{I}-\mathbf{n}_a\mathbf{n}_a) ] =0,
 \quad
 \mathbf{v}^s_2\cdot\mathbf{n}_a=0,
\end{equation}
where $\mathbf{n}_a$ is a unit normal to the axis of symmetry in the
$(r,z)$-plane. At the point in the $(r,z)$-plane where the
(initially hemispherical) free surface meets the syringe tip we have
a pinned contact line:
\begin{equation}\label{pinned_shape}
 f(1,1,t)=0 \qquad (t\ge0).
\end{equation}

It is assumed that in the far field, the exterior fluid and the liquid
inside the syringe are at rest whilst on the cylinder's surface, the no-slip condition is applied
to both phases, so that
\begin{equation}\label{pinned_shape}
\mathbf{u},~\mathbf{u}_g\to\mathbf{0} \qquad\hbox{as}\qquad
r^2+z^2\to\infty,\qquad\qquad  \mathbf{u}=\mathbf{u}_g=\mathbf{0} \qquad\hbox{at}\quad r=1,z\ge1.
\end{equation}

As the initial conditions, we set that both the liquid and the fluid
are at rest, the free surface is in equilibrium
\begin{equation}\label{ini-rest}
 \mathbf{u}=\mathbf{u}_g=\mathbf{0},\qquad  \rho^s_1=\rho^s_{1e},
 \qquad(t=0),
\end{equation}
and the free surface has the shape of a hemisphere
\begin{equation}\label{ini-shape}
 f(r,z,0)=r^2+(z-1)^2-1=0.
\end{equation}

For computations using the conventional model, the formulation described above can still be used if the parameter $\epsilon$ is set to zero so that the interface formation dynamics is `turned-off'.

\section{Computational details}

In order to solve the problem, we employ the finite-element-based
computational platform described in \cite{sprittles_ijnmf,sprittles_jcp}, where one can find a
user-friendly step-by-step algorithm to its implementation \footnote{See the Appendix of this paper for a small correction to \cite{sprittles_jcp}}. In the
present work, we only need (a) to extend it to incorporate the
dynamics of the ambient fluid, which can be done in a straightforward
way and (b) adjust the problem formulation described above for the
numerical treatment. The latter means, firstly, truncating the
computational domain by introducing the `far-field' boundary at a
large but finite distance from the origin. This far-field boundary
is shown schematically in Fig.~\ref{F:sketch}, where $r_{far}$ and
$z_{far}$ have to be sufficiently far away from the origin for their
location and the soft boundary conditions we impose there to have a
negligible effect on the coalescence dynamics.

The second adjustment that we have to make is to introduce a small
but finite radius $r_{min}$ of the initial contact of the two drops,
so that initially we have the internal interface at $0<r<r_{min}$,
$z=0$, where we will use the initial condition
\begin{equation}\label{ini-intl}
\rho^s_2=\rho^s_{1e},
\end{equation}
stating that the trapped part of the free surface has not yet
started relaxing towards its eventual equilibrium state of
$\rho^s_2=1$ (at which point the surface has no tension $\sigma_2=0$). As the initial shape of the free surface we will take
simply a truncated sphere/hemisphere satisfying $z(r_{min})=0$.
\begin{equation}\label{ic1}
 (r-r_{min})^2+(z-z_0)^2 = z_0^2,
 \end{equation}
where $z_0=\hbox{$\frac{1}{2}$}(1+(1-r_{min})^2)$, so that, if
there is no base, i.e.\ $r_{min}=0$, one has $z_0=1$ and hence
$r^2+(z-1)^2=1$, thus recovering (\ref{ini-shape}).

If the conventional model is used, as described in detail in \cite{sprittles_pof2}, the initial free-surface shape is taken from the analytic solution in \cite{hopper84}, obtained for Stokes flow, to be
\begin{eqnarray} \notag
 r(\theta) = \sqrt{2}\left[(1-m^2)(1+m^2)^{-1/2}
 (1+2m\cos\left(2\theta\right)+m^2)^{-1}\right](1+m)\cos\theta, \\  \label{ic2}
 z(\theta) =
 \sqrt{2}\left[(1-m^2)(1+m^2)^{-1/2}(1+2m\cos\left(2\theta\right)+m^2)^{-1}\right]
 (1-m)\sin\theta,
\end{eqnarray}
for $0<\theta<\theta_u$, where $m$ is chosen such that $r(0)=r_{min}$
is the initial bridge radius, which we choose, and $\theta_u$ is
chosen such that $r(\theta_u)=z(\theta_u)=1$. Notably, for
$r_{min}\to0$ we have $m\to1$  and $r^2+(z-1)^2=1$, i.e.\ the drop's
profile is a semicircle of unit radius which touches the plane of
symmetry at the origin as required.

Importantly, unlike the conventional model, for the interface formation model the
limit $r_{min}\to0$ does not give rise to a singularity
\citep{shik07}, so that here a non-zero value of $r_{min}$ is used
for convenience of the computations and no special shape is required to artificially enforce smoothness of the free surface. We will look at how the value of $r_{min}$ influences
the outcome of computations in \S\ref{S:ini}.

\section{Simplifications for $Q, \bar{\alpha}\to 0$ with $A=O(1)$}\label{S:simpler}

For the case of an inviscid dynamically-passive ambient fluid the full
model, set out in \S\ref{S:problem}, has been studied in
\cite{sprittles_pof2}. The results of this study suggest an
asymptotic simplification that facilitates the computations. Before
extending the earlier work to include the full dynamics of the
viscous ambient gas or liquid, we take the limit $\bar{\alpha}\to0$ with $A=O(1)$. As shown by
experiments \citep{shik97a,blake02}, $\alpha\sim\mu^{-1}$ and, given that
$\bar{\alpha}=\alpha \mu/R$, for the class of liquids considered in
the present work, $\alpha\mu\approx10^{-9}$~m \citep{blake02} so
that $\bar{\alpha}=10^{-9}\hbox{[m]}/R\hbox{[m]}$ and, for drops
that are larger than a micron, $R>10^{-6}$~m, we have
$\bar{\alpha}<10^{-3}\ll1$. Importantly, unlike the case of dynamic
wetting (\cite{shik07}), the solution to the coalescence problem
remains singularity-free in the limit $\bar{\alpha}\to0$, $A=O(1)$,
so that, to leading order, we can simply set $\bar{\alpha}=0$ in the
above formulation.

The computations in the framework of the simplified system have been
compared to results for the full system of equations as computed in
\cite{sprittles_pof2}.  In the range of parameters considered in
\cite{sprittles_pof2}, and therefore those considered here, curves
for all the relevant quantities proved to be very close for the two systems and, consequently, henceforth
the simpler system will be used.  It is important to note, however,
that as much smaller scales are approached, such a simplification may no
longer be valid.

The essence of what this asymptotic limit is about is very simple.
From (\ref{tang-bulk_vel-free}) and (\ref{Darcy-free}), it can be
seen that, in this limit, the differences between the components of
velocity tangential to the free surface on either side of the
surface as well as between these components and the surface velocity itself become negligible,
$\mathbf{u}_{\parallel}=\mathbf{u}_{g\parallel}=\mathbf{v}^s_{1\parallel}$.
(Here subscripts $\parallel$ denotes the component of a vector
tangential to a surface, i.e.\ it represents the convolution of the
vector with the tensor $(\mathbf{I}-\mathbf{n}\mathbf{n})$ which
extracts the tangential components of vectors and tensors.) In other
words, in this limit we recover the classical condition of
continuity of the tangential component of velocity across an
interface. Similarly, on the internal interface, from
(\ref{Darcy-intl}), we have that
$\mathbf{u}_{\parallel}=\mathbf{v}^s_{2\parallel}$. Basically, the
limit $\bar{\alpha}\to0$, $A=O(1)$ in our system of equations leads
to the transport of surface mass along the interfaces being due to
the bulk velocity tangential to that interface, rather than by
surface tension gradients acting inside the interface, i.e.\ to the
situation one has in the classical fluid-mechanics model.

When extending this approach from an inviscid dynamically passive gas to the case of a viscous gas, as with the conventional model, the effect of the gas now manifests itself only through the balance of stress terms (\ref{norm-stress}) and (\ref{tang-stress-free}), with $\mathbf{P}_g$ composed of a dynamic pressure and non-zero viscous terms.

A further simplification is to also consider $Q\rightarrow 0$ in (\ref{continuity-free}), that is to assume that the flux of mass into/out of the interface affects only the surface dynamics, rather than the bulk flow also.  In other words, as in the classical case, the normal velocity is also now continuous across an interface so that on the free surface $\mathbf{u}\cdot\mathbf{n}=\mathbf{u}_g\cdot\mathbf{n}=\mathbf{v}^s_1\cdot\mathbf{n}$ and at the plane of symmetry $\mathbf{u}\cdot\mathbf{n}_s=\mathbf{v}^s_2\cdot\mathbf{n}_s=0$.  Given that $Q=\rho^s_{(0)}/(\rho\sigma\tau_{\mu})$, estimates suggest that $Q\sim10^{-2}\ll1$ for the liquids considered so that the effect of taking $Q=0$ is also negligible.  Simulations with a finite $Q$ have confirmed this.  As a result, one has the conventional formulation used for capillary flows in which the surface tension on an interface is considered dynamic combined with an equation of state (\ref{eqstate}) and the surface continuity equation:
\begin{equation}\label{sifm}
 \epsilon\left[\frac{\partial\rho^s_i}{\partial t}
 +\nabla\cdot\left(\rho^s_i\mathbf{u}\right)\right]
 =-\left(\rho^s_i-\rho^s_{ie}\right),\qquad i=1,2,
\end{equation}
where $\rho^{s}_{2e}=1$. As the single surface equation only contains first-order derivatives in space, equation (\ref{cl1}) is no longer applicable and a single boundary condition is applied on the surface density where each interface meets the axis of symmetry or the syringe tip.  In contrast with the case of dynamic wetting, where the motion of the contact line relative to the solid forces the mass flux through the contact line, in the case of coalescence the contact line moves as if to minimise the stress in its vicinity and does so by becoming a stagnation line for the bulk flow (in the reference frame moving with the contact line). As a result, condition (\ref{cl1}) appears to be satisfied with both terms on the left-hand side equal to zero.

In taking the aforementioned simplifications, the interface formation model is stripped down to its simplest form which has the advantage that (a) there are less free parameters that have to be estimated and (b) it becomes easier to isolate the key features of the interface formation model that distinguish it from the conventional model's predictions.  A notable consequence of the simplified formulation is that the ambient fluid can either be a viscous gas or immiscible liquid, in contrast to the case of $Q\neq 0$ where a second liquid would mean that the mass exchange on both sides of the interface would have to be considered \citep{shik07}.

\section{Parametric study of the model}

To establish an appropriate parameter range, and to compare to experimental data from \cite{paulsen11} in \S\ref{S:exp}, consider the parameter values based on water-glycerol drops, and initially take them to have $R=2$~mm radii. These mixtures have the advantage that their surface tension with air $\sigma=65$~mN~m${}^{-1}$ and density $\rho=1200$~kg~m${}^{-3}$ remain approximately constant, whilst the viscosity can range over three orders of magnitude $\mu=10^{-3}$--$1$~Pa~s.  Then the Reynolds number is in the range $Re=10^{-1}$--$10^5$.  For coalescence in air of density $\rho_g=1.2$~kg~m${}^{-3}$ and viscosity $\mu_g=18$~$\mu$Pa~s, the gas-to-liquid density ratio is $\bar{\rho} = 10^{-3}$ throughout and the
viscosity ratio will be in the range $\bar{\mu} = 10^{-5}$--$10^{-2}$.

Since our problem is both nonlinear and multi-parametric, a sensible
strategy for exploring it would be to use the estimates for the material constants of the interface
formation model obtained from experiments on dynamic wetting \citep{blake02}, where the model was used without any alterations, as a `base case' and to investigate how a variation of these parameters influences the model's predictions. The experimental results in \cite{blake02} suggest that for the class of fluids
considered, i.e.\ water glycerol mixtures, the relaxation time of
the interface $\tau = \tau_{\mu}\mu$, where
$\tau_{\mu}$ is approximately constant across all the mixtures
considered, so that the dimensionless parameters are $\epsilon =
\sigma\tau_{\mu}/R$, $\rho^s_{1e} =
(\rho^s_{1e})_{dim}/\rho^s_{(0)}$ and $\lambda =
\gamma\rho^s_{(0)}/\sigma_{1e}$. Using previous estimates for these
parameters as our base state (denoted with a subscript `0'), about which the parameters can be varied to identify their role, we have
\begin{align}\label{base_parameters}
\epsilon_0 =3.3\times10^{-5} ,\quad (\rho^s_{1e})_0=0.4,\quad \lambda_0 = (1-\rho^s_{1e})^{-1},
\end{align}
where $\tau_{\mu}=10^{-6}$~m$^2$~N$^{-1}$, consistent with previous estimates in \cite{blake02}, so that the relaxation time for a viscosity of $\mu=10^{-3}$~Pa~s (water) is $1$~ns whilst for $\mu=1$~Pa~s (roughly, pure glycerol) it is $\tau=1~\mu$s.  In our comparison to experiment in \S\ref{S:exp}, values from (\ref{base_parameters}) will be used across all liquids, in contrast to \cite{sprittles_pof2} where $\rho^s_{1e}$ was fitted.  At present, no study has been conducted the influence of this parameter on the mixture's properties, so, again, the simplest possibility is considered.

In order to further our understanding of the mechanisms governing the
interface formation/disappearance process, we will look into how
varying these parameters around their base state influences the
propagation of the liquid bridge connecting the coalescing drops for
the intermediate viscosity of $\mu=48$~mPa~s, so that $Re_0=68$ and $\bar{\mu}_0=4\times10^{-4}$.

\subsection{Influence of initial conditions}\label{S:ini}

\begin{figure}
     \centering
     \includegraphics[scale=0.4]{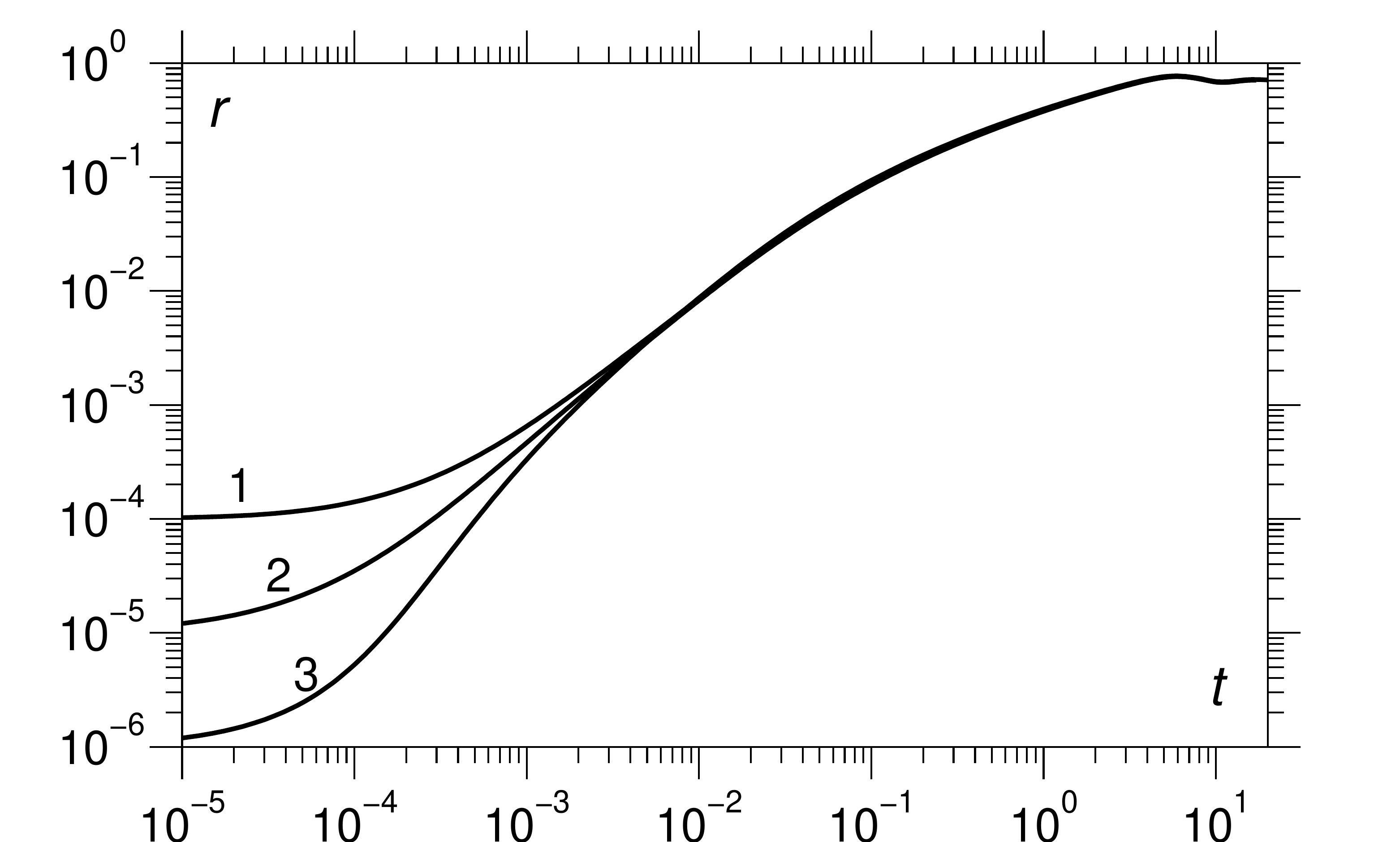}
 \caption{Effect of decreasing initial bridge radius for the base case ($Re_0=68$ and $\bar{\mu}_0=4\times10^{-4}$), with 1: $r_{min}=10^{-4}$ to 2: $r_{min}=10^{-5}$ and 3: $r_{min}=10^{-6}$.  As expected, the value of $r_{min}$ becomes insignificant shortly after the start of the process.}
 \label{F:rmin}
\end{figure}

Before considering the effect of the interface formation parameters on the coalescence event, we would like to establish the effect which our initial conditions, in particular the finite bridge radius $r_{min}$ which the computations are started at, have on the subsequent dynamics.  In Figure~\ref{F:rmin} the effect of the initial radius $r_{min}$ in (\ref{ic1}) is shown, and it can be seen that after a certain time, or distance, all the curves fall on top of each other. Specifically, one can see that after around $r=10r_{min}$ the effect of the initial conditions has diminished and the curves begin to fall onto a single line.  A similar result has been obtained for the conventional model.  Thus, henceforth, computational results will be shown from $r=10r_{min}$ so that the range under consideration has not been affected by the finite initial radius, i.e.\ the same curve would be obtained for smaller $r_{min}$.   This reinforces the point that it is not the amount of trapped interface that is initially formed that matters, but the subsequent dynamics.

Notably, although with the interface formation model $r_{min}=10^{-5}$ can easily be resolved, so that $r>10^{-4}$ becomes independent of $r_{min}$, these values cannot be achieved with computations of the conventional
model, where the radius of curvature at the bridge front becomes
prohibitively small for $r_{min}<10^{-4}$, due to the requirement that, in the framework of this model, the bridge must be smooth. As a result, although in the parametric study we look at $r>10^{-4}$, in order to ensure both the conventional and interface formation models are treated on an equal footing in our comparison with experiments in \S\ref{S:exp}, computed curves and experimental results are considered from $r=10^{-3}$ following a limitation imposed by the conventional model.

\subsection{Role of the interface formation parameters}\label{S:role}

\begin{figure}
     \centering
     \includegraphics[scale=0.34]{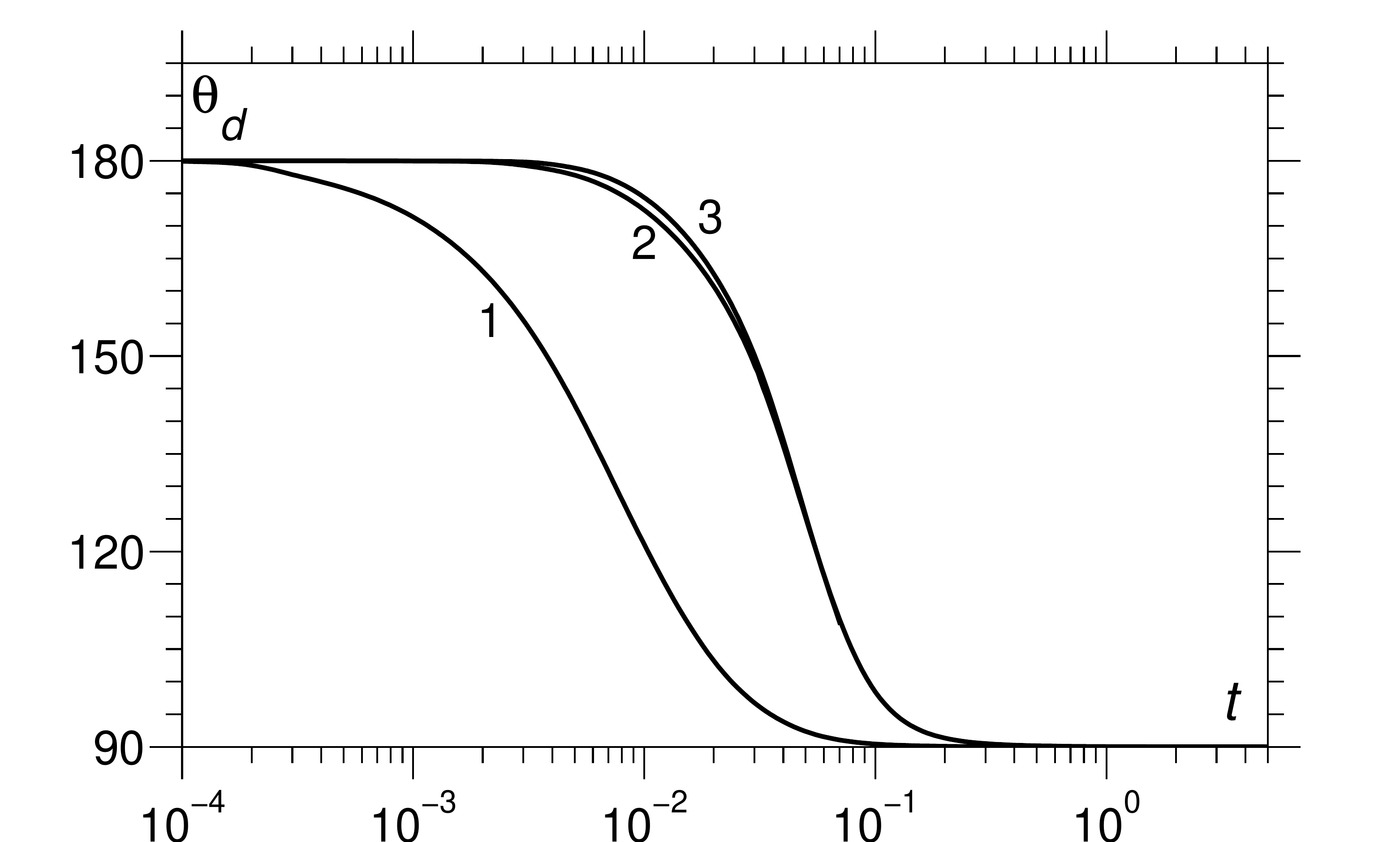}   
 \caption{Influence of interface formation parameters on the evolution of the contact angle $\theta_d$ for $Re_0=68$ and $\bar{\mu}_0=4\times10^{-4}$. Curve~1 is the base case used in previous calculations, curve~2 is for $10\epsilon_0$, curve~3 is for $\lambda = 10\lambda_0$. The last two curves highlight the existence of three distinct stages of the process: (a) the very initial stage where the drops `touch' with their free surface forming a cusp, (b) the relaxation stage where there is an evolving corner between free surfaces and (c) the `equilibrium' stage when the conventional model takes over.}
\label{F:48cP_angle}
\end{figure}

\begin{figure}
     \centering
    \includegraphics[scale=0.4]{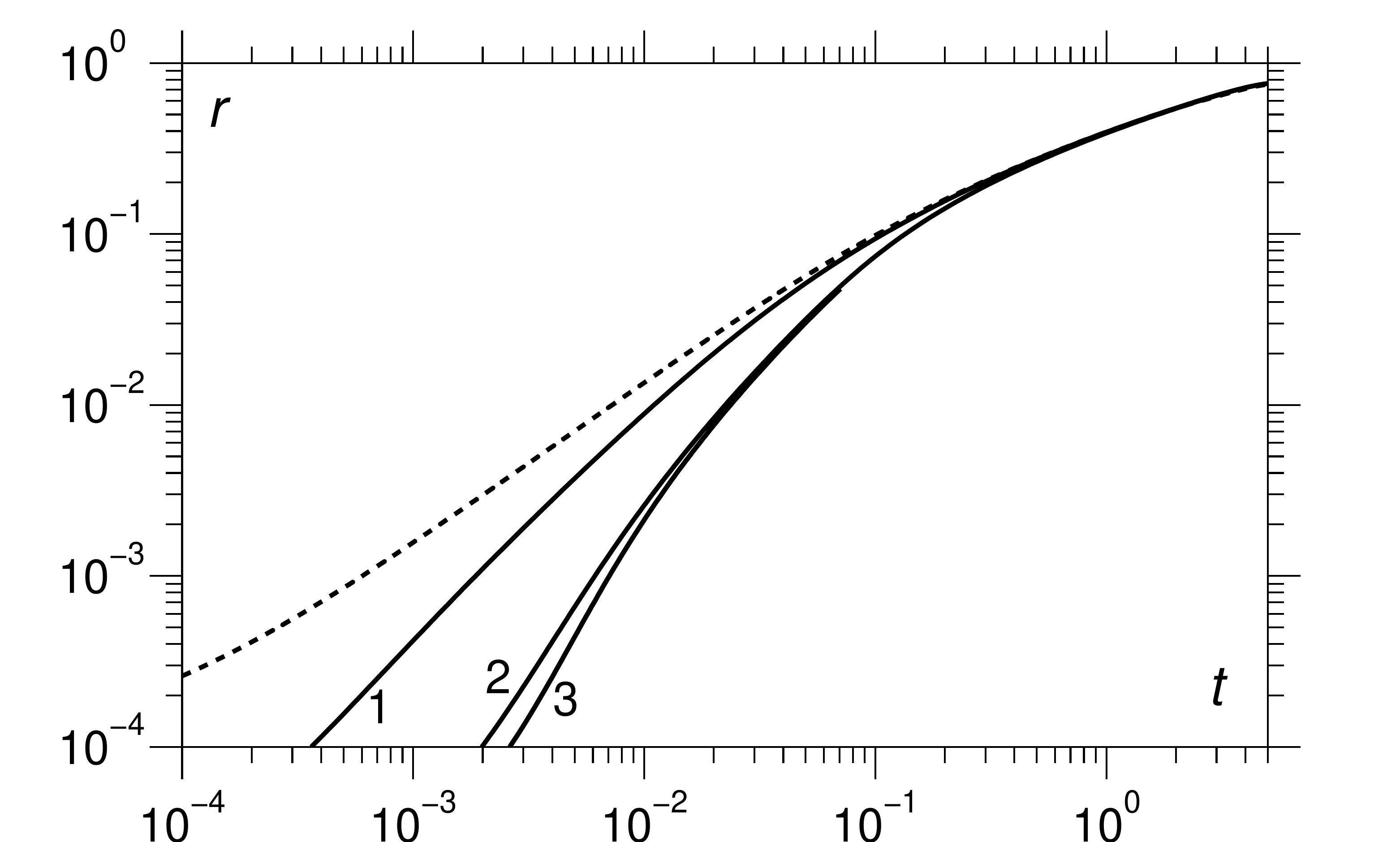}    
 \caption{Influence of interface formation parameters at $Re_0=68$ and $\bar{\mu}_0=4\times10^{-4}$. Curve~1 is the base case used in previous calculations, curve~2 is for $10\epsilon_0$, curve~3 is for $\lambda = 10\lambda_0$. The dashed line is the conventional model's prediction.}
\label{F:48cP_ana}
\end{figure}

From Figure~\ref{F:48cP_angle}, one can see that the stage during which the interface formation dynamics is occurring, i.e.\ the period in which the free surface is not smooth ($\theta_d\neq90^{\circ}$),  comprises of different regimes.  In
what we will refer to as the very initial stages, around
$t<T_{ini}=10^{-3}$ for the base state (curve~1), the angle at which
the free surface meets the plane of symmetry stays approximately constant
$\theta_d\approx180^\circ$.  This can be seen most clearly in curves 2 and 3.  Then one has the `relaxation stage',
around $T_{ini}<t<T_{rel}=10^{-2}$ for the base state, where the angle
relaxes to $\theta_d\approx90^\circ$ after which
the interface formation model turns into the conventional one, as can be
seen in Figure~\ref{F:48cP_ana}.
\begin{figure}
     \centering
    \includegraphics[scale=0.4]{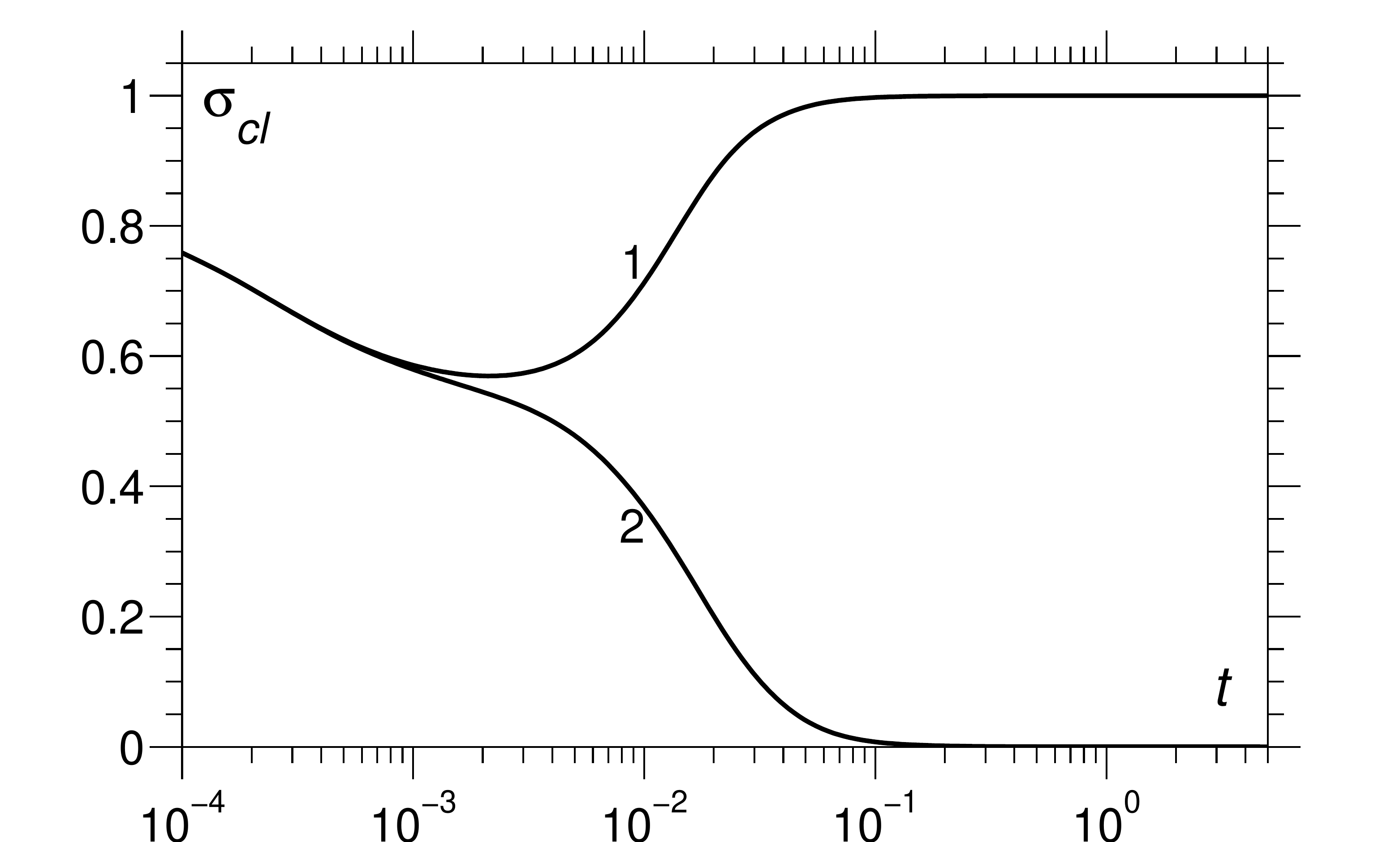}
 \caption{Evolution of the surface tension where the free surface meets the plane of symmetry (at the `contact line') on 1: the free surface and 2: the internal interface obtained for $Re_0=68$ and $\bar{\mu}_0=4\times10^{-4}$.  The very initial stage where the two surface tensions evolve together due to the stretching of both interface by the bulk flow is followed by the relaxation stage where they evolve towards their equilibrium values.  Once these values are reached the conventional model takes over.}
\label{F:48cP_sigma}
    \end{figure}

Figure~\ref{F:48cP_sigma} shows the time
dependence of the surface tension at the contact line $\sigma_{cl}$
on each interface.  As we can see, in the very initial stage
$\sigma_{1}=\sigma_{2}=\sigma_{ini}(t)$, evolving from $\sigma_{ini}(t)=1$ at
$t=0$ imposed by our initial conditions to
$\sigma_{ini}(t)\approx0.6$ at $t=T_{ini}=10^{-3}$. This is a very interesting feature as it indicates that both the free surface and the internal interface are being stretched, approximately in equal measure, as the process of coalescence goes through its very initial stage. From the force
balance at the contact line (\ref{cl2}), $\sigma_{1}=\sigma_{2}$
leads immediately to $\theta_d=180^\circ$, as already noted from
Figure~\ref{F:48cP_angle}.

At time $t=T_{ini}$, the relaxation
stage takes over, during which the liquid-fluid and internal interface approach their
equilibrium states of $\sigma_{1}=1$ and $\sigma_{2}=0$ at
$t=T_{rel}$.  For the liquid-fluid interface, which started in
equilibrium but was then driven out of this state by the coalescence
process, this relaxation stage involves an increase in surface tension, whilst for
the internal interface, the surface tension continues to decrease
until it reaches a state where this interface has effectively
`disappeared' and no longer has surface properties which would distinguish
it from the bulk phase.  In other words, the process of coalescence
as modelled by the interface formation model is complete when
$t=T_{rel}$.  Given that the process naturally divides into these
different stages, we now consider the effect of the interfacial
parameters in each as illustrated in Figure~\ref{F:48cP_ana}.

The two parameters which we vary, $\epsilon$ and $\lambda=1/(1-\rho^s_{1e})$, can be seen from Figure~\ref{F:48cP_angle} to affect the time during which $\theta_d\approx180^\circ$, i.e.\ the time scale of the very initial stages $t<T_{ini}$.  Roughly, a factor of ten increase in either $\epsilon$ (curve 2) or $\lambda$ (curve 3) is seen to increase the time of the very initial stage by a factor ten. From Figure~\ref{F:48cP_ana}, we can see that the longer the coalescence process spends in this initial stage, the slower the initial motion.  

The two controlling parameters are seen to have a similar affect on the relaxation stage, with this period extending by a factor of ten to $T_{rel}\approx10^{-1}$, as opposed to $T_{rel}\approx10^{-2}$ for the base case, when either $\epsilon$ or $\lambda$ are increased by a factor of ten.  Figure~\ref{F:48cP_ana} confirms what one may expect, that the earlier the interface formation process is over the faster the bridge speed propagation will be. This is to be expected, as in the limit $T_{rel}\rightarrow 0$, in which $\theta_d=90^\circ$ in an infinitesimal time, we have the dynamics of the conventional model which is known to have a singular velocity at the start of the process, so that larger $T_{rel}$ must give a slower coalescence speed.  Notably, we see that for $t>0.1$, all the curves coincide as the conventional model takes over so that the interfacial parameters no longer have an influence on the dynamics.

\subsection{Influence of the ambient fluid}\label{S:gas}

To consider the influence of the ambient fluid on the coalescence process, all the base parameters remain fixed with the exception of $\bar{\mu}$ and $\bar{\rho}$ which are now allowed to vary.  For $\bar{\rho}\leq 0.01$, corresponding to the range of realistic liquid-gas systems, which are our main focus here, the influence of the finite gas density on the dynamics of coalescence is seen to be negligible so that, henceforth, we will consider only the effect of the viscosity ratio $\bar{\mu}$.

\begin{figure}
     \centering
     \includegraphics[scale=0.4]{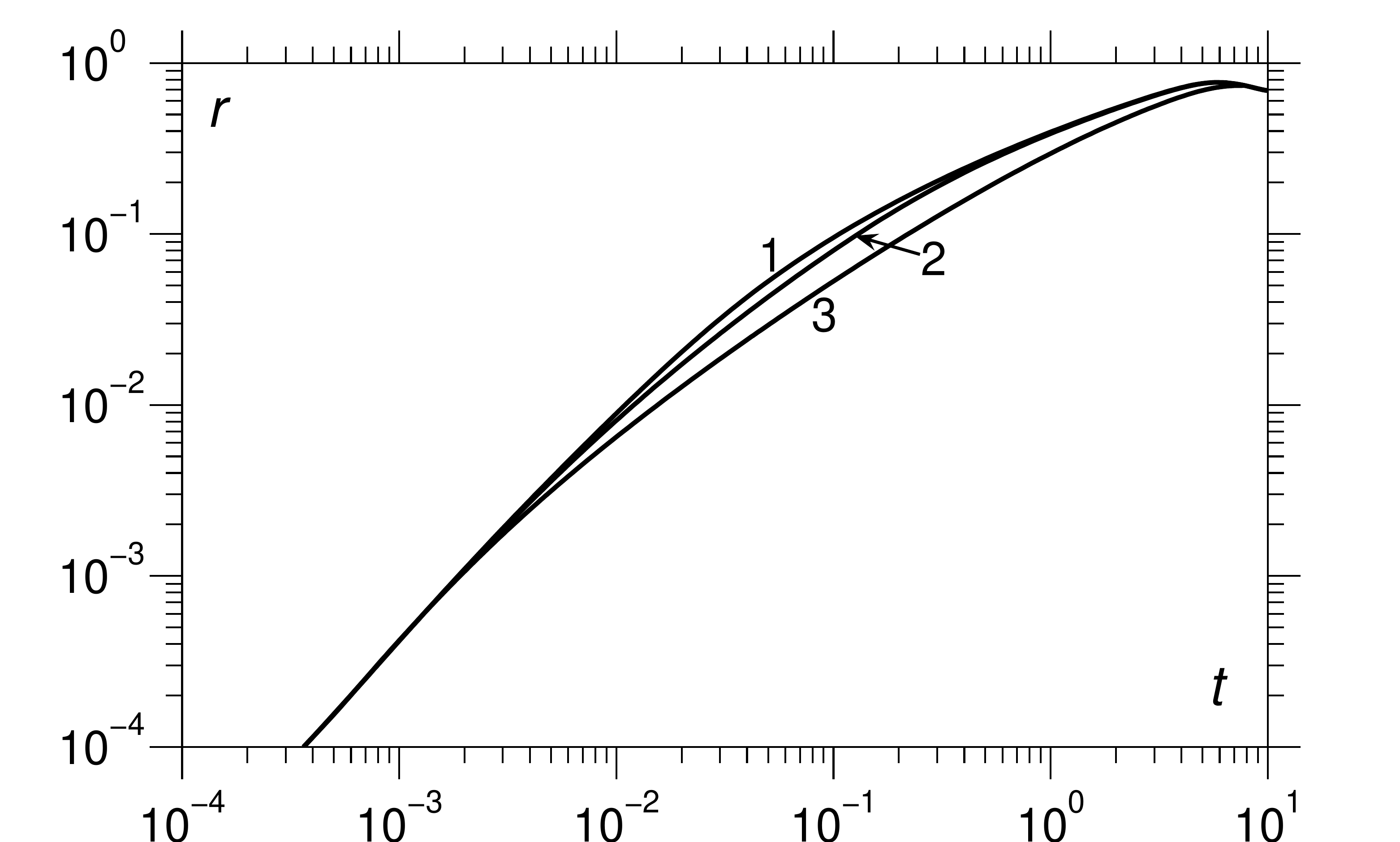}
 \caption{Effect of the viscosity ratio on bridge propagation, with all other parameters set to their base values. Curves 1: $\bar{\mu}=10^{-4}$, 2: $\bar{\mu}=10^{-2}$ and 3: $\bar{\mu}=1$. Strikingly, although the gap between the free surfaces is narrowest in the very initial stage of the process, the ambient fluid-to-liquid viscosity ratio has no effect there.}\label{F:mubar}
\end{figure}

In Figure~\ref{F:mubar}, the effect of the viscosity ratio on the bridge propagation is shown and what is immediately striking is that for $r<10^{-2}$ an increase of two orders of magnitude in viscosity ratio, from $\bar{\mu}=10^{-4}$ (curve 1) to $\bar{\mu}=10^{-2}$ (curve 2), has little effect on the speed of coalescence.  Similarly, increasing by four orders of magnitude to $\bar{\mu}=1$ does not alter the bridge evolution for $r<10^{-3}$.  Only after a finite time does the initial indifference to $\bar{\mu}$ give way to an effect of viscosity ratio: as one would expect, larger viscosity ratios result in a slower speed of coalescence. Comparing $\bar{\mu}=10^{-4}$ (curve 1) and $\bar{\mu}=10^{-2}$ (curve 2), it can be seen that only in the period $10^{-2}<r<10^{-1}$ is there a noticeable difference caused by the change in $\bar{\mu}$.

\begin{figure}
     \centering
\includegraphics[scale=0.4]{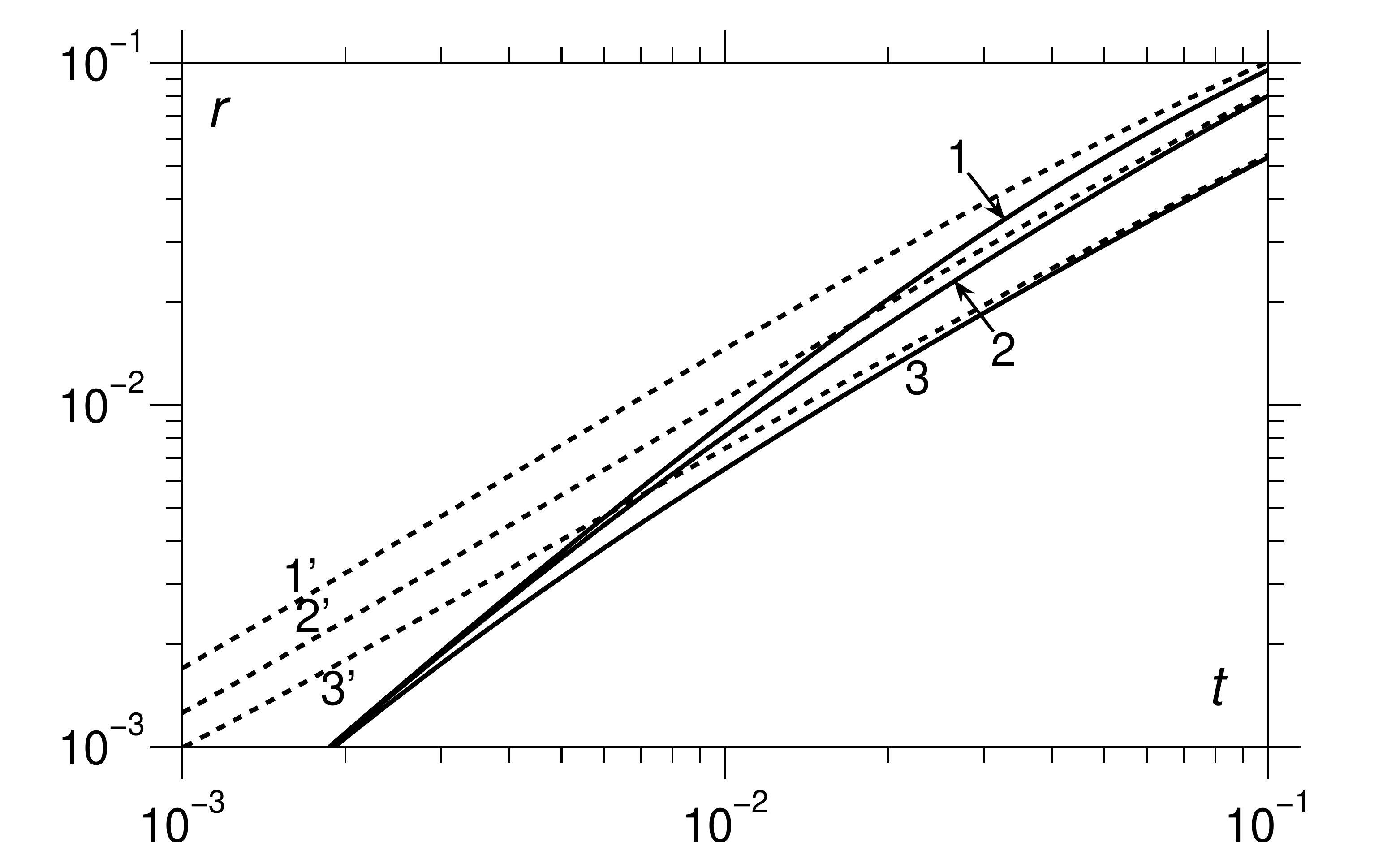}
 \caption{Effect of the viscosity ratio on bridge propagation for the two different models, with all other parameters set to their base values. Curves 1: $\bar{\mu}=10^{-4}$, 2: $\bar{\mu}=10^{-2}$ and 3: $\bar{\mu}=1$.  The predictions of the conventional model are shown by dashed lines, with curve numbers labelled with a prime.}\label{F:mubar_both}
\end{figure}

Figure~\ref{F:mubar_both} shows that the behaviour observed for the interface formation model (solid lines) differs significantly from that computed for the conventional model (dashed lines).  In the conventional model, the effect of the viscosity ratio is instantaneously felt and the deviation caused by the differences in $\bar{\mu}$ remains roughly constant (on a log-log plot) for $r<0.1$.  In contrast, for the interface formation model, only after a finite time do the curves fan-out, with the distance between them increasing whilst $r<0.1$.

\begin{figure}
     \centering
     \includegraphics[scale=0.4]{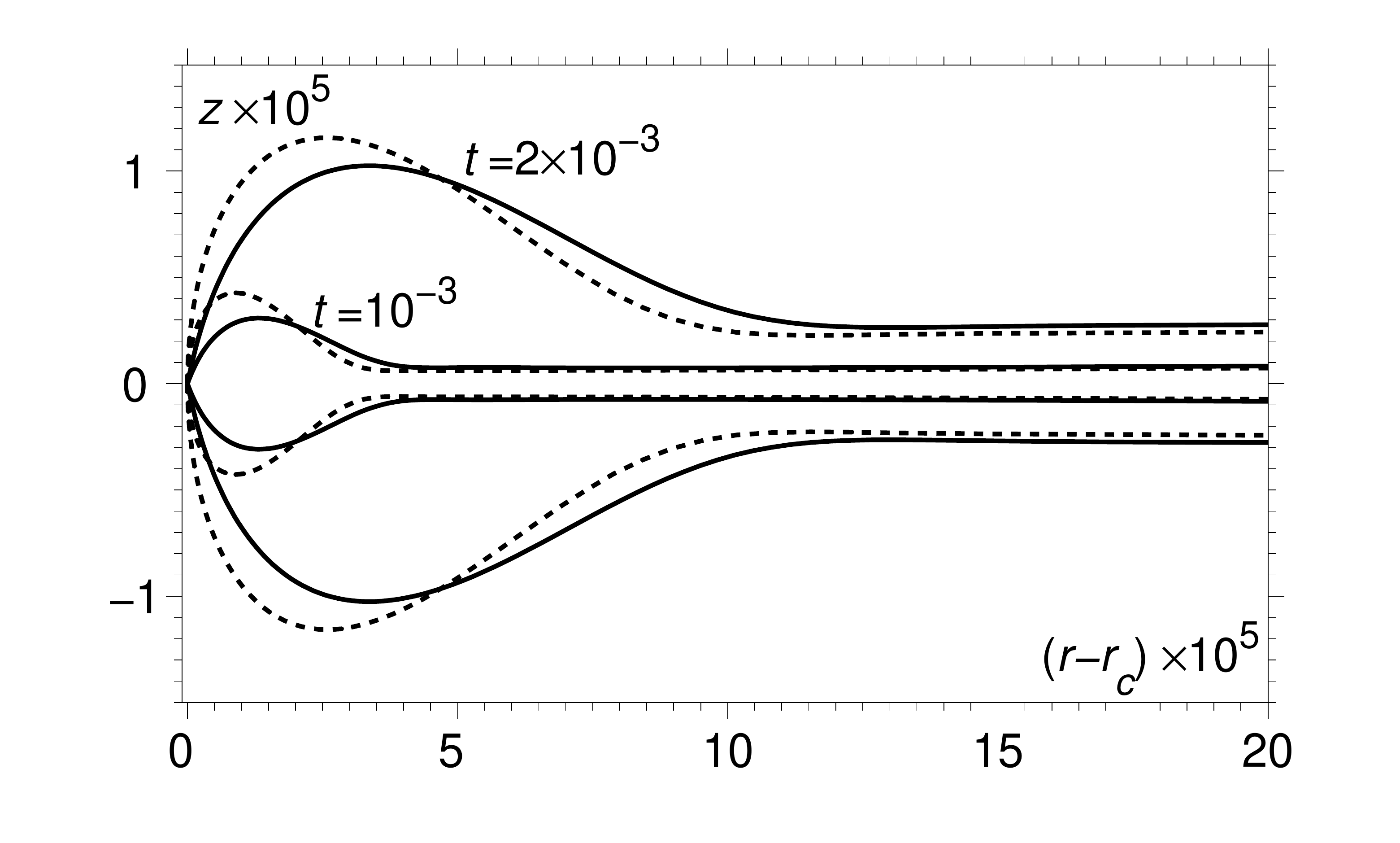}
 \caption{Free surface shapes obtained from the interface formation model (solid lines) and the conventional model (dashed lines) at $t=10^{-3}$ and $t=2\times10^{-3}$.}\label{F:fs_shape}
\end{figure}

In an attempt to understand the observed behaviour, in Figure~\ref{F:fs_shape}, the free surface shape of the bridge front is shown at two instances in time $t=10^{-3}, 2\times10^{-3}$ for the two different models.  It can be seen that for the interface formation model at $t=10^{-3}$, the free surface is not smooth and compared to the conventional model's shape at this time there is less of a `bubble' of fluid trapped in front of the bridge.  It could be that it is this geometrical feature, present only in the interface formation dynamics, which results in the ambient fluid having less of an effect for this model than for the conventional model where the free surface is always smooth (dashed lines).  In other words, as the process enters the relaxation stage and the angle evolves from $\theta_d=180^\circ$ to $\theta_d=90^\circ$, ambient fluid is more easily swept away from the bridge front region and thus has little effect on the dynamics.  In contrast, in the conventional model a bubble of fluid builds up in front of the bridge front so that its dynamics and removal become necessary for the bridge to propagate and thus its behaviour, governed by the value of $\bar{\mu}$, alters the speed of coalescence.  For the interface formation model, once the free surface becomes smooth and matters are handed over to the conventional one, similar effects are observed.

\subsection{Effect of drop size}\label{S:size}

\begin{figure}
     \centering
     \includegraphics[scale=0.4]{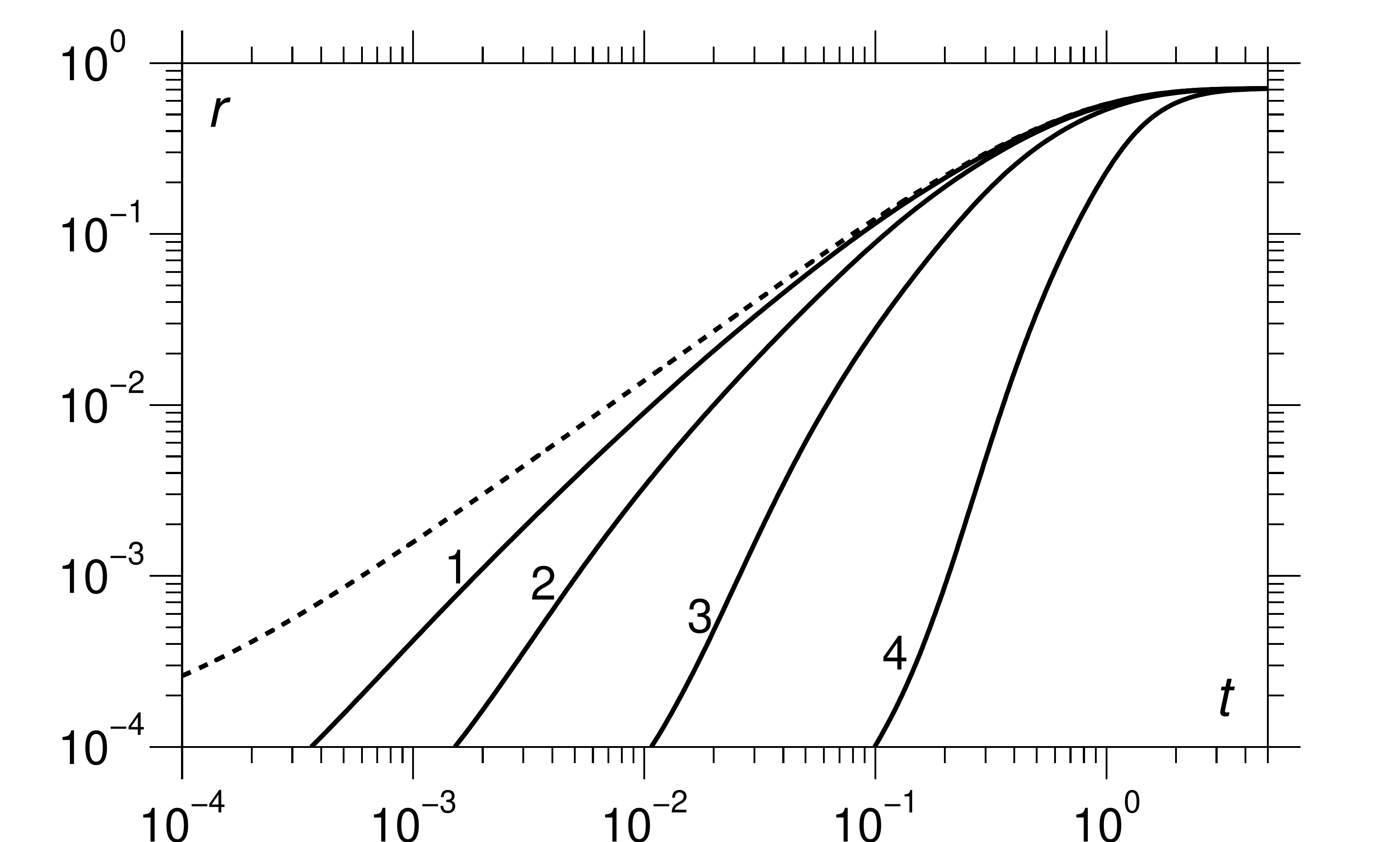}
     \includegraphics[scale=0.4]{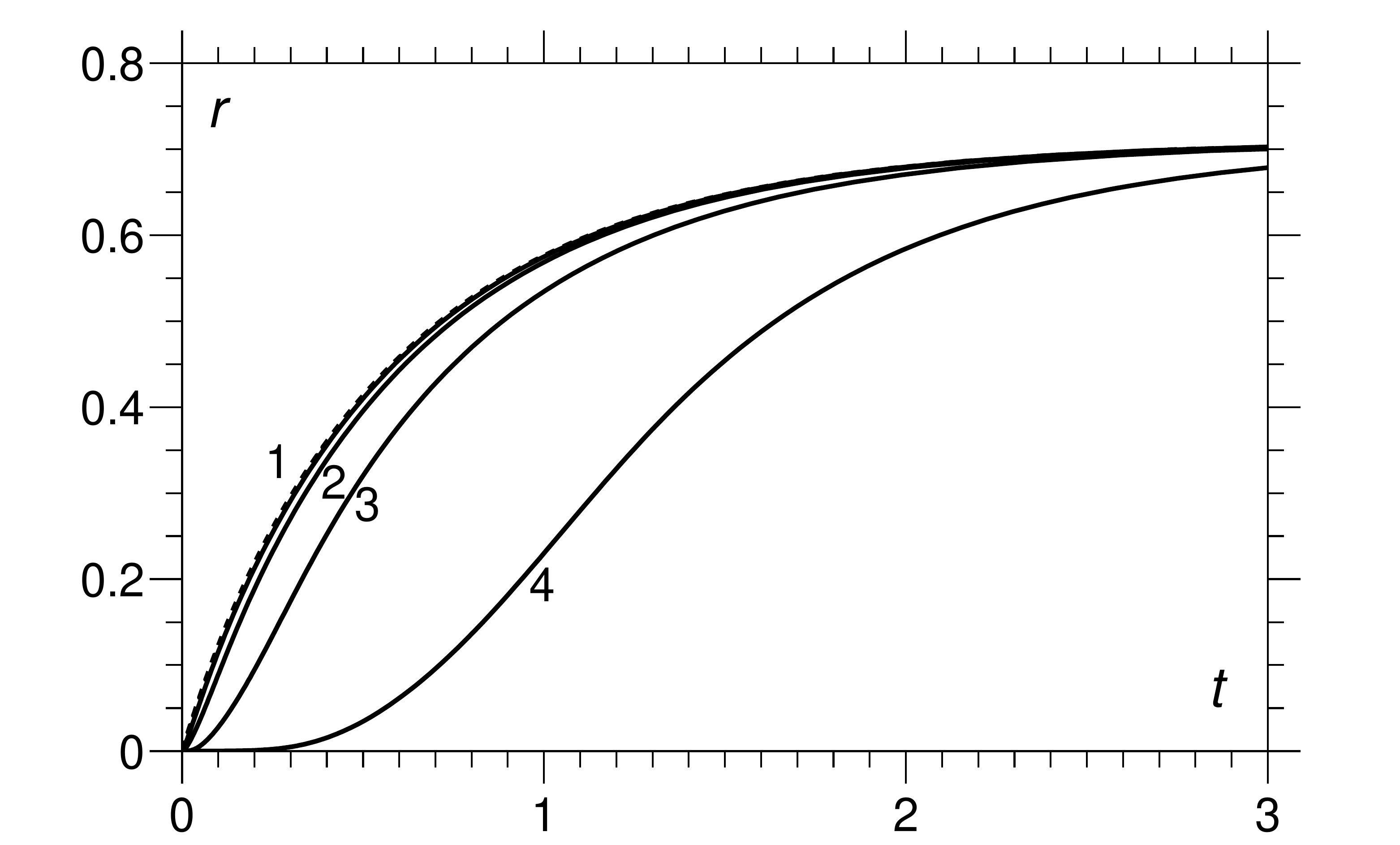}
 \caption{Affect of decreasing the radius of the drops on the bridge propagation with 1: $R=2$~mm, 2: $R=200~\mu$m, 3: $R=20~\mu$m and 4: $R=2~\mu$m.  The dashed line is the computed solution for the conventional model, which for the $Re=0$ case considered here is independent of drop size.}
 \label{F:size}
\end{figure}

Consider how the speed of coalescence depends on the size of the drop $R$. To further simplify matters consider the usual base case except for $Re=0$, i.e.\ a high viscosity solution.  In this case, there is a universal curve describing the bridge propagation for the conventional model, this curve is shown as a dashed line in Figure~\ref{F:size}.  However, for the interface formation model the parameter $\epsilon = \sigma\tau_{\mu}/R = 6.5\times10^{-8} R^{-1}$ depends on the size of the drop.  Therefore, the conventional model predicts no change in the curve relating (dimensionless) bridge radius $r$ to (dimensionless) time $t$, whilst the interface formation model could predict some effect, and it is this which will now be quantified.

In Figure~\ref{F:size}, the bridge evolution is shown on both log-log and linear plots. From an experimental perspective it is likely that the linear plot will prove most useful, so it is this we shall focus on.  What can be seen is that although differences can be observed on the log-log plot for larger drops $R>200~\mu$m (curves 1,2), deviations from the conventional model's predictions (dashed line) are relatively small on the linear one.  However, for $R=20~\mu$m (curve 3) a noticeable deviation from the conventional model's universal curve is observed and for $R=2~\mu$m a huge change is seen with the coalescence speed dramatically reduced.

\begin{figure}
     \centering
     \includegraphics[scale=0.4]{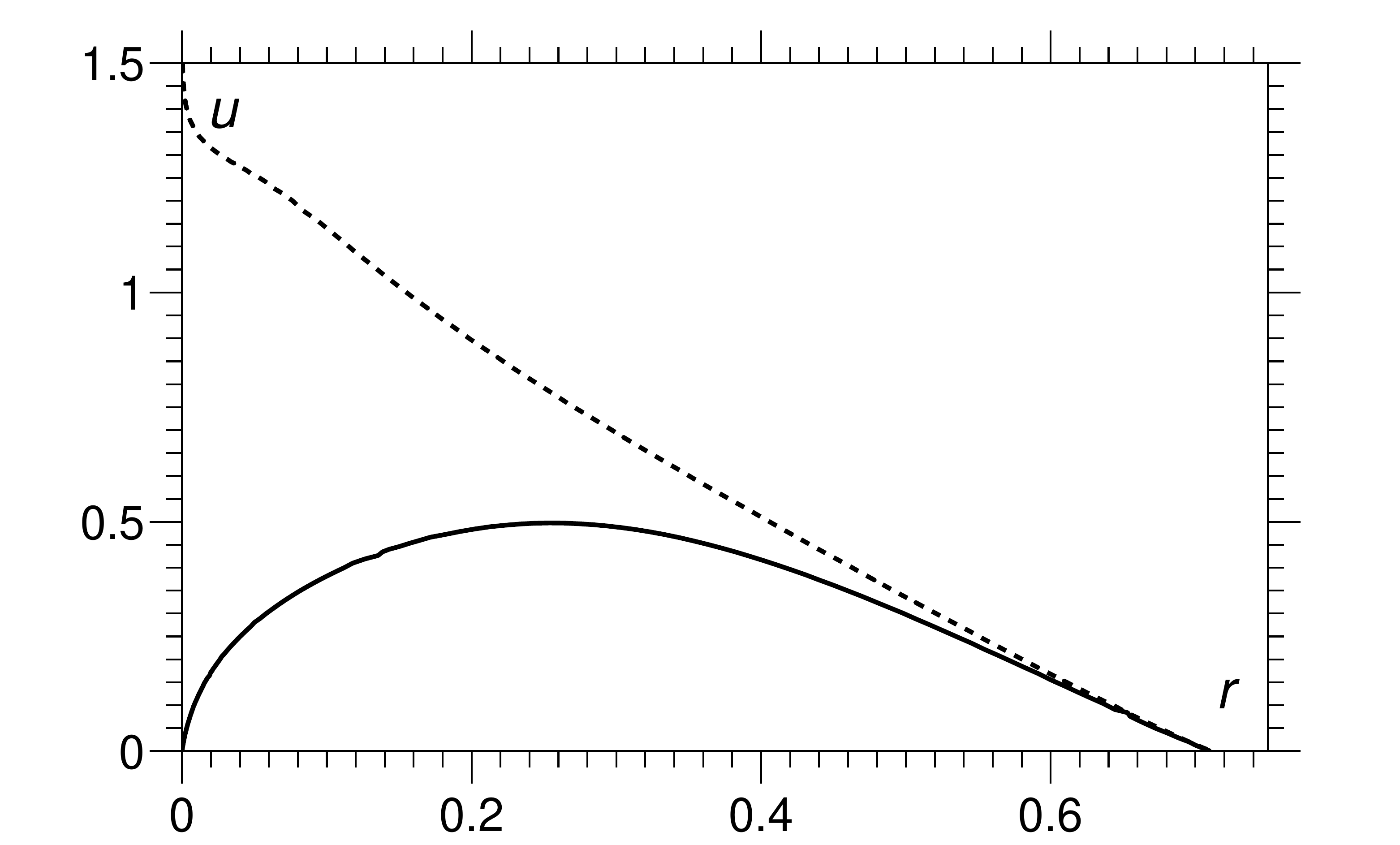}
 \caption{Bridge speed $u$ as a function of bridge radius for $R=2~\mu$m with the solid line corresponding to the interface formation model's predictions and the dashed line the conventional model's}
 \label{F:speed}
\end{figure}

The changes for small drops are most easily seen in Figure~\ref{F:speed}, where the bridge speed for the smallest drops considered ($R=2~\mu$m) is plotted as a function of the bridge's radius.  The conventional model's universal solution predicts a monotonically decreasing speed  throughout the coalescence process whilst the interface formation model predicts a clear maximum around $r=0.25$. From an experimental perspective, it may be easier to observe this maximum in the bridge speed for small drops, rather than trying to compare the bridge evolution across different sized drops as shown in Figure~\ref{F:size}.

Figure~\ref{F:speed} highlights the key difference between the two models' predictions:  the conventional model is singular, whilst the interface formation model is singularity-free.  Specifically, in the case of the interface formation model, in Figure~\ref{F:speed} we can see that the initial speed of coalescence is very small, and only over a finite and relatively large time does the bridge front accelerate to a maximum, before relaxing towards its equilibrium (static) shape.  In contrast, the initial speed for the conventional model is singularly large and it can only go down as free-surface curvature and hence the capillary pressure that drives the pressure decreases.  Although such features are present in all calculations, it is only when the drop's size becomes comparable to the scales on which the interface formation physics acts that these effects visibly change the global motion of the drops.

\section{Comparison with experiments}\label{S:exp}

As our base parameters have been setup in order to align with the experiments in \cite{paulsen11}, all that is required here is to specify the viscosity of the particular mixtures we will consider, which are chosen to give the widest possible range of parameters, and these are $\mu=3.3,~48,~230$~mPa~s.  Then, the Reynolds numbers are
$Re=1.4\times10^4,~68,~2.9$ and for coalescence in air of density
$\rho_g=1.2$~kg~m${}^{-3}$ and viscosity $\mu_g=18$~$\mu$Pa~s, the
gas-to-liquid density ratio is $\bar{\rho} = 10^{-3}$ and the
viscosity ratios are, respectively, $\bar{\mu} =
5.5\times10^{-3},~3.8\times10^{-4},~7.8\times10^{-5}$.

At this stage, we could look to fit our two parameters for the interface formation model to the experimental data.  In particular, in \cite{sprittles_pof2} better agreement between theoretical predictions and experiments was obtained by varying $\rho^s_{1e}$ as a function of fluid viscosity.  It is perfectly reasonable for such a variation to occur as the nature of the interface changes with the percentage of glycerol in the mixture; however, here we continue our approach of considering only the simplest possible model and thus use the same parameters across all viscosities.

\begin{figure}
     \centering
     \includegraphics[scale=0.4]{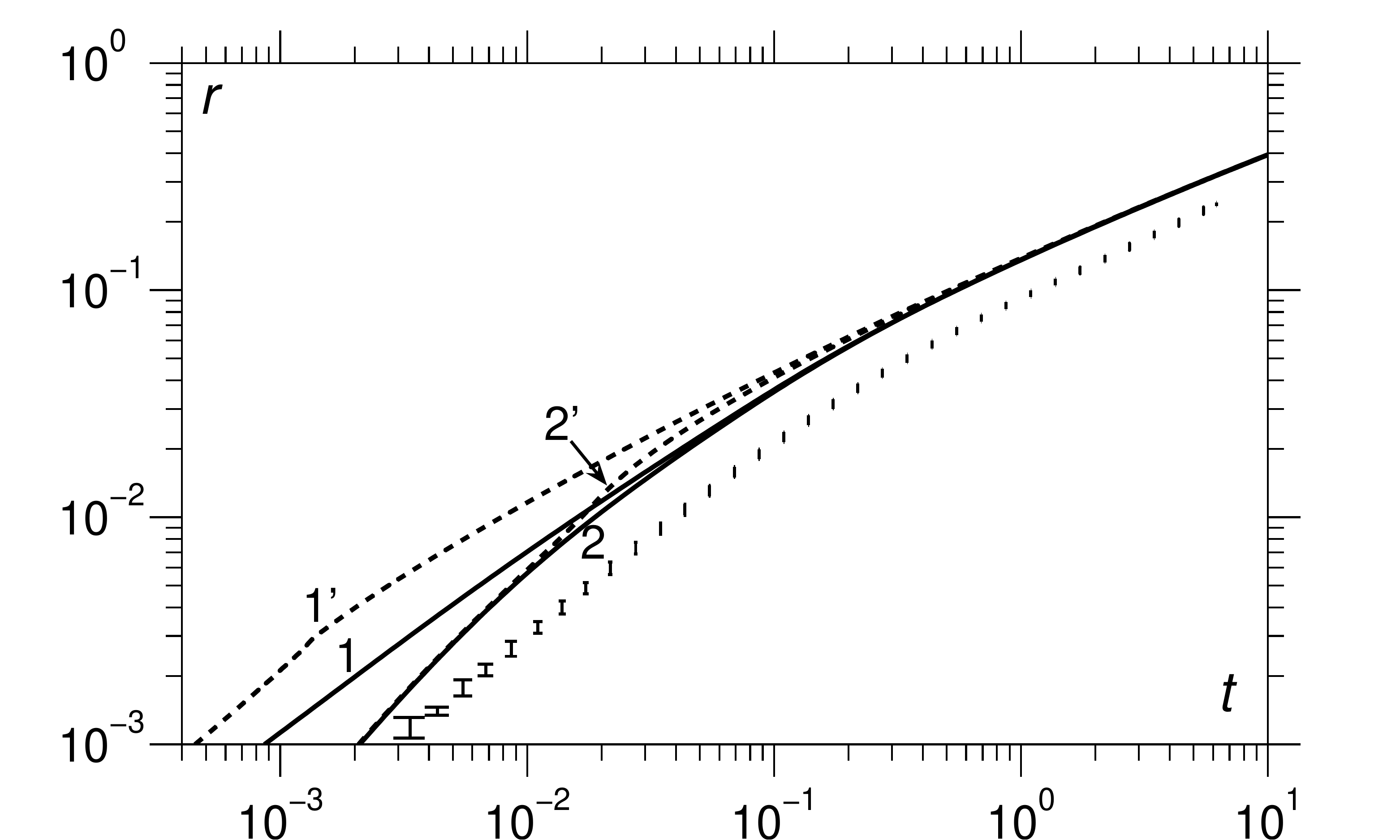}
\includegraphics[scale=0.4]{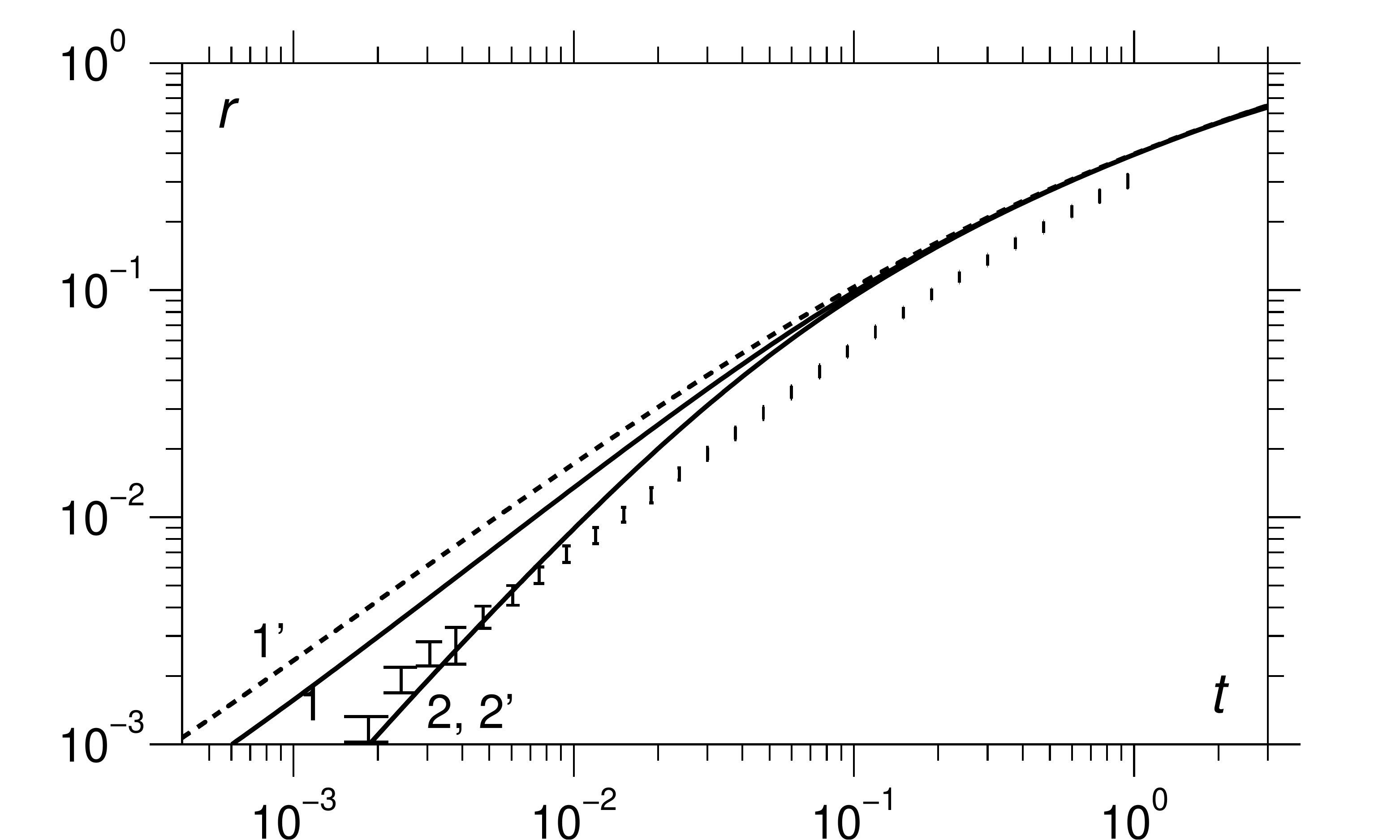}
\includegraphics[scale=0.4]{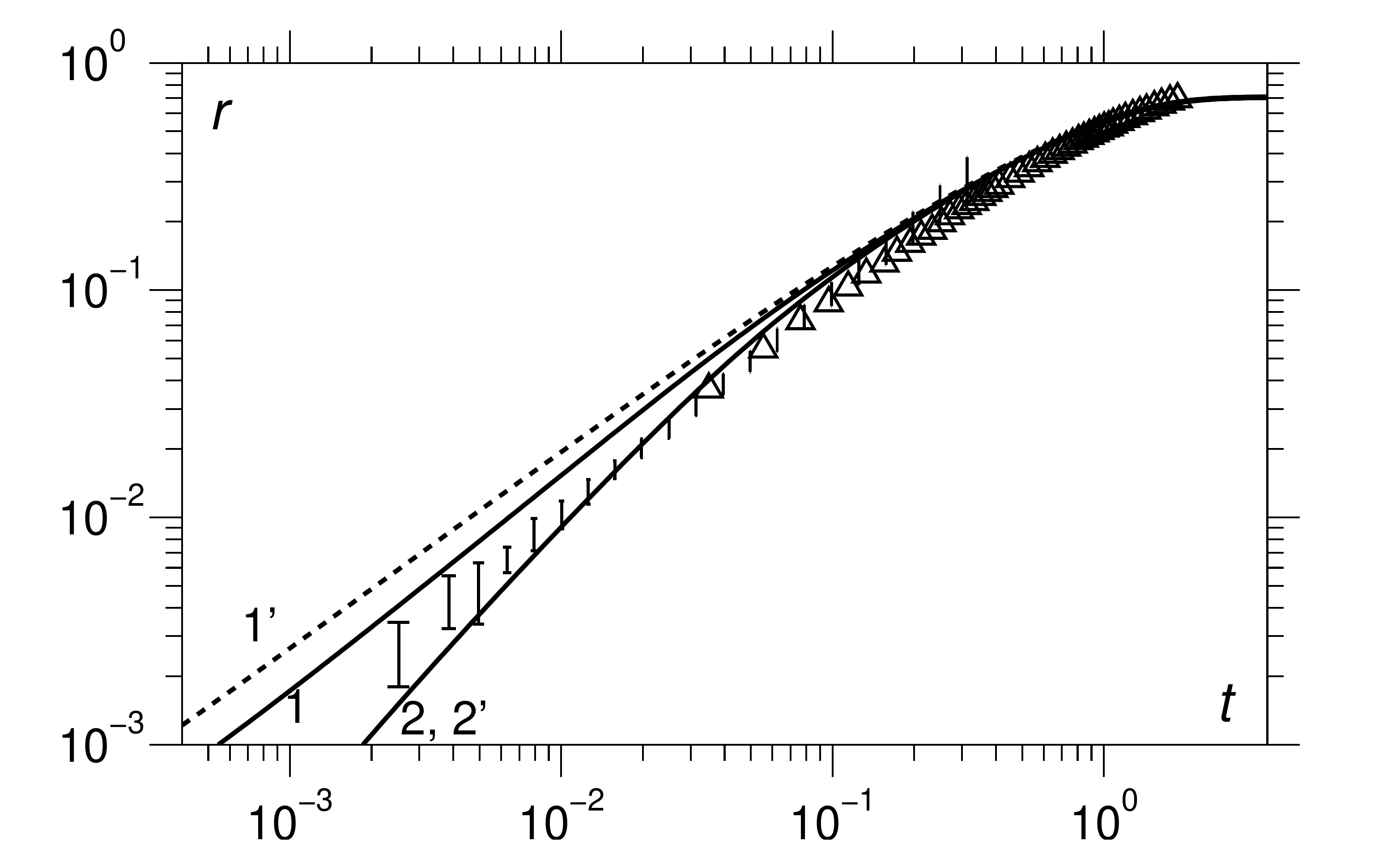}
 \caption{Comparison of the predictions of (a) the conventional model
 (curve~1) and (b) the interface formation model (curve~2) with
 experiments from \cite{paulsen11} (error bars) and \cite{thoroddsen05} (triangles), for $Re=1.4\times10^4,~68,~2.9$.  Curves marked with a prime are those computed for each model when the gas is considered passive.}
 \label{F:ifm}
\end{figure}

In Figure~\ref{F:ifm}, curves obtained from both the conventional model (curves 1) and the interface formation model (curves 2) are shown for both the case in which the surrounding air is considered viscous and for the situation where the gas is assumed passive (dashed curves marked with a prime).  It is apparent that in all cases the viscosity of the gas influences the conventional model's curves, whilst it is only at the lowest viscosity that the gas has a noticeable effect on the interface formation model's predictions. This is consistent with our findings in \S\ref{S:gas}, where we saw that for computations with the interface formation model there is only a small window during which the gas can have an effect on the coalescence speed, which only occurs here at the lowest liquid viscosity, whereas in the conventional model a finite viscosity always has an effect.

In terms of the actual agreement between the models' predictions and the experiments, the viscous gas does not alter the conclusions of \cite{sprittles_pof2}: the
interface formation model gives a better description of the initial stages of coalescence across two orders of magnitude change in liquid viscosity than the conventional model, whilst for low viscosity mixtures both models deviate from the experimental measurements of the \emph{later} stages of coalescence.  The possibility that this discrepancy between theory and experiments is caused by the effects of gravity, interface formation and/or the ambient fluid have all been ruled out, suggesting progress in uncovering the reason can only be made by conducting further theory-driven experiments.

\section{Discussion}

The computational simulations performed have highlighted the role of
the ambient fluid's dynamics in the coalescence process and have shown that the effect is different for the two models considered. Despite
this, the conclusions from \cite{sprittles_pof2} remain unchanged.
In particular, both the conventional model and the interface
formation model, when the latter reduces to the former, give similar predictions for the final stages of the
coalescence process, roughly on the mm-scale, where optical
measurements are available. For the flow on the microscale, singularities inherent in the
conventional formulation lead to an overprediction of the speed of
coalescence.  It is only due to the recent experimental
results in \cite{paulsen11} that the errors in the conventional
model's predictions could be brought to light.  In contrast, the interface
formation model is singularity-free in the initial stages and describes
the experimental data better, even with the simplifying assumptions used to reduce the number of free parameters.

An analysis of the evolution of the interface formation process sheds
light on the time scales involved during a coalescence event. Of
particular note is that the time scales recovered are much larger
than the relaxation time of the interface, which
would be an obvious initial estimate for these scales.  The cause of
this phenomenon has not been fully accounted for, but it appears to
be related to the unsteady nature of the process, with rapid
variations in the shape of the interfaces, combined with the initial
far-from-equilibrium configuration of the system. When combined,
these effects sustain the non-equilibrium interfacial dynamics.
What would be of particular interest is the development of an
asymptotic theory for the different stages of the process which may
shed additional light on how the interface is maintained in its
non-equilibrium state.  Furthermore, such a theory, or scaling law,
could make a comparison of the interface formation theory with
experimental data a more routine task, rather than requiring full
computation at every stage.

So far, we have focussed on the dynamics of the bridge of a mm-sized
drop, where we saw that the results obtained in the framework of
different models for the bridge dynamics on the microscale differ
significantly, but these differences do not affect the global
dynamics of the drops as the interface formation-disappearance
processes are over long before the global dynamics comes into play. In contrast, in \S\ref{S:size} we saw that for a
micro-drop the global dynamics of the drop is heavily dependent on the
model used.  Consequently, theory-driven experiments in this range can target
global features of the drop coalescence process such as, say, the
aspect ratio of the drop, as often used to characterise oscillating
drops \citep{sprittles_pof}.  In this way, by studying smaller
drops, optical measurement again becomes a viable method for probing
the physics of the coalescence process.

\section*{Acknowledgements}

The authors would like to thank Dr J.D.~Paulsen, Dr J.C.~Burton and Professor S.R.~Nagel for providing us with the data from their experiments published in \cite{paulsen11} and Dr Y. Li for spotting the typo in \cite{sprittles_jcp}.

\section*{Appendix: Correction to \cite{sprittles_jcp}}

Since publishing our user-friendly step-by-step guide to the finite element implementation of the interface formation model in the Appendix of \cite{sprittles_jcp}, a typo has been brought to our attention which is present in the text, but not in the code which has been developed. In particular, equation (53) should read:
\begin{equation}
\nabla^s\cdot\mathbf{a}^s_{||} = \pdiff{a^s_t}{s} + \frac{n a^s_{t.r}}{r}, \qquad a^s_{t.r}=(\mathbf{a}^s\cdot\mathbf{t})(\mathbf{t}\cdot\mathbf{e}_r)
 \end{equation}
and, consequently, the second term on the right-hand-side of equation (63), should be changed from
\begin{equation}
n\rho^s_{\gamma,j}\diff{r_{\gamma,k}}{t}\int_{s_{\gamma e}}\phi_{\gamma,i}\phi_{\gamma,j}\phi_{\gamma,k}ds_{\gamma,e} \qquad\hbox{to}\qquad
n\rho^s_{\gamma,j}c^s_{t,k}\int_{s_{\gamma e}}\phi_{\gamma,i}\phi_{\gamma,j}\phi_{\gamma,k}t_r ds_{\gamma,e}.
\end{equation}
It is important to stress that the correct equations were always used in our code.

\bibliographystyle{jfm}
\bibliography{Bibliography}

\end{document}